\documentclass[11pt]{article}
\pdfoutput=1
\usepackage{amsmath}
\usepackage{amssymb}
\usepackage{graphicx,bbm,mathrsfs}
\usepackage{nicefrac}
\usepackage{slashed}
\usepackage{bbm}
\usepackage{geometry}
\usepackage{empheq}

\usepackage{jheppub}

\usepackage{mathtools}

\usepackage{dcolumn}   
\usepackage{bm}        
\usepackage{graphicx,mathrsfs}
\usepackage{nicefrac}
\usepackage{multirow}
\usepackage{color}
\usepackage{mathtools}
\usepackage{makecell}

\usepackage{environ} 
\usepackage{lipsum} 
 \NewEnviron{Smaller11}{
           \scalebox{1.1}{$\BODY$} 
 } 
 \NewEnviron{Smaller08}{
           \scalebox{0.8}{$\BODY$} 
 }

\newcommand{\be}{\begin{equation}}
\newcommand{\ee}{\end{equation}}
\newcommand{\bea}{\begin{eqnarray}}
\newcommand{\eea}{\end{eqnarray}}

\newcommand{\n}{\mathbf{n}}
\newcommand{\J}{J}
\newcommand{\D}{{\cal D}}
\newcommand{\Lv}{{\bf L}}
\newcommand{\F}{{ \bm F}}
\newcommand{\x}{{\bm x}}
\newcommand{\xeps}{x_\epsilon}

\newcommand{\bmu}{{\bm \mu}}
\newcommand{\bpartial}{{\bm \partial}}
\newcommand{\W}{W}

\newcommand{\DeltaD}{\Delta}

\newcommand{\q}{k}

\renewcommand{\S}{\mathcal{S}}

\usepackage{wasysym}
\usepackage{hhline,colortbl}
\usepackage{xcolor}

\newcommand{\nn}{\nonumber}

\mathchardef\mhyphen="2D

\usepackage{titlesec}

\titleformat*{\section}{\Large\bfseries}
\titleformat*{\subsection}{\large\bfseries}
\titleformat*{\subsubsection}{\large\bfseries}
\titleformat*{\paragraph}{\large\bfseries}
\titleformat*{\subparagraph}{\large\bfseries}

\makeatletter
\newcommand*{\prodsym}{%
  \DOTSB
  \mathop{
    \mathchoice
      {\rlap{\kern.3em\rotatebox[origin=c]{-90}{}}{\prod}}
      {\vcenter{\rlap{\kern.2em\rotatebox[origin=c]{-90}{}}}{\prod}}
      {\sum}{\sum}
  }\slimits@
}
\makeatother

\makeatletter
\DeclareFontFamily{OMX}{MnSymbolE}{}
\DeclareSymbolFont{MnLargeSymbols}{OMX}{MnSymbolE}{m}{n}
\SetSymbolFont{MnLargeSymbols}{bold}{OMX}{MnSymbolE}{b}{n}
\DeclareFontShape{OMX}{MnSymbolE}{m}{n}{
    <-6>  MnSymbolE5
   <6-7>  MnSymbolE6
   <7-8>  MnSymbolE7
   <8-9>  MnSymbolE8
   <9-10> MnSymbolE9
  <10-12> MnSymbolE10
  <12->   MnSymbolE12
}{}
\DeclareFontShape{OMX}{MnSymbolE}{b}{n}{
    <-6>  MnSymbolE-Bold5
   <6-7>  MnSymbolE-Bold6
   <7-8>  MnSymbolE-Bold7
   <8-9>  MnSymbolE-Bold8
   <9-10> MnSymbolE-Bold9
  <10-12> MnSymbolE-Bold10
  <12->   MnSymbolE-Bold12
}{}

\let\llangle\@undefined
\let\rrangle\@undefined
\DeclareMathDelimiter{\llangle}{\mathopen}%
                     {MnLargeSymbols}{'164}{MnLargeSymbols}{'164}
\DeclareMathDelimiter{\rrangle}{\mathclose}%
                     {MnLargeSymbols}{'171}{MnLargeSymbols}{'171}
\makeatother

\begin{document}

\vspace*{4mm}

\begin{center}

\thispagestyle{empty}
{\Huge
Scalar-Mediated  Quantum Forces Between Macroscopic Bodies and Interferometry
 }\\[12mm]

\renewcommand{\thefootnote}{\fnsymbol{footnote}}

{\large  
Philippe~Brax~$^a$\footnote{phbrax@gmail.com} and Sylvain~Fichet~$^{b,c}$\footnote{sfichet@caltech.edu }\,
}\\[8mm]

\end{center}

\noindent \quad\quad\quad $^a$ \textit{Institut de Physique Th\'{e}orique, Universit\'e Paris-Saclay, CEA, CNRS, }

\noindent \quad\quad\quad\quad \textit{ F-91191 Gif/Yvette Cedex, France}
\\

\noindent  \quad\quad\quad $^{b}$\textit{ ICTP South American Institute for Fundamental Research  \& IFT-UNESP,}

\noindent \quad\quad\quad\quad \textit{R. Dr. Bento Teobaldo Ferraz 271, S\~ao Paulo, Brazil}
\\ 

\noindent \quad\quad\quad $^{c}$\textit{ Centro de Ciencias Naturais e Humanas, Universidade Federal do ABC,}

\noindent \quad\quad\quad\quad \textit{Santo Andre, 09210-580 SP, Brazil}

\addtocounter{footnote}{-1}

\vspace*{12mm}

\begin{center}
{  \bf  Abstract }
\end{center}

We study the  quantum force between  classical objects mediated by massive scalar fields  bilinearly coupled to matter. 
The existence of such fields is motivated by dark matter, dark energy, and by the possibility of a hidden sector beyond the Standard Model. 
We introduce the   quantum work felt by an arbitrary (either rigid or deformable) classical body in the presence of the scalar and show that it is finite  upon requiring conservation of matter. As an example,  we  explicitly  show that  the quantum pressure inside a Dirichlet  sphere is finite --- up to renormalizable divergences.
With this method we compute the scalar-induced quantum  force in simple planar geometries.
In  plane-point geometry we show how to compute the contribution of the  quantum force to  the   phase shift  observable in  atom interferometers. 
We show that atom interferometry is likely to become a competitive search method for light particles bilinearly coupled to matter, provided that the interferometer arms have lengths below $\sim 10$\,cm.

\newpage
\setcounter{tocdepth}{2}
\tableofcontents

\newpage

\section{Introduction}
\label{se:intro}


 The exchange of virtual particles  induces macroscopic forces between bodies. Beyond the tree level,
such forces are  relativistic and quantum in essence, they are properly described in the framework of quantum field theory (QFT).  Here we refer to such forces simply  as ``quantum forces''. 
Even though the seminal works on Casimir forces are from the nineteen forties \cite{CP_original,Casimir_original},  the topic of  quantum forces  is still very much active   (see \textit{e.g.}  \cite{Milton:2004ya, Klimchitskaya:2009cw,2015AnP...527...45R, Woods:2015pla, Bimonte:2017bir, Bimonte:2021maf,Bimonte:2022een} for recent { {papers}} and  reviews).
In the case of electromagnetic interactions, refined calculations take into account  medium properties  such as electromagnetic permittivities as  well as effects from  finite temperature, see \cite{bordag2009advances} for a comprehensive review.  In the present paper we work  in the simple framework of scalar-mediated quantum forces at zero temperature, with the key assumption that the scalar couples bilinearly  to the sources.\,\footnote{See \textit{e.g.} \cite{Milton:2002vm,Graham:2002fw,Jaffe:2005vp,Mobassem:2014jma} for a selection of related  works involving  scalar quantum forces.}
In this scalar setting, we use a simple variational method  to derive quantum forces between bodies of various shapes and positions.

The scalar case is of prime relevance for cosmology \cite{Hui:2016ltb,Joyce:2014kja} and deriving quantum forces mediated by massive scalars could lead to new  laboratory tests of important cosmological models, from scalar dark matter to dark energy. In this paper, we argue that atom interferometry \cite{Hamilton:2015zga} could be such a promising technique.
 Our  primary motivation comes from the pervasiveness  of scalar fields  bilinearly coupled to matter in cosmology \cite{Joyce:2014kja} and in extensions of the Standard Model of particle physics \cite{Allanach:2016yth}. 
For instance, the dark matter in our Universe   can be modelled as a scalar particle with $Z_2$ symmetry, which couples thus bilinearly  to matter.
A wealth of dark energy models and related modified gravity models  \cite{Brax:2021wcv}  also involve a light scalar field. For these, when the classical force  induced by the linear coupling  to sources  is screened  (see \textit{e.g.} \cite{Damour:1994zq,Khoury:2003aq,Khoury:2003rn}), the quantum force arising from the bilinear coupling  to sources can become dominant \cite{Brax:2018grq}, hence motivating our study.   Finally, irrespective of  observational motivations, the existence of a  sector hidden beyond the Standard Model and featuring a light scalar with a $Z_2$ symmetry (\textit{e.g.} a pseudo-Nambu Goldstone boson of a symmetry of the hidden sector) is  a logical possibility that requires investigation. 
There is thus, in short, a cornucopia of reasons to study scalar fields bilinearly coupled to matter.

A secondary motivation is that some of the technical and conceptual results that we present --- such as the proof of finiteness of the quantum work ---  are best exposed with a scalar field. 
We use an approach based on the variation of the quantum vacuum energy which is similar in spirit  to the one found in the seminal work by Schwinger~\cite{Schwinger:1977pa}. 
The developments presented here  can  be taken as a streamlined presentation of this variational  method based in the context of  scalar QFT.

The vacuum energy of QFT features divergences which are treated via standard renormalization methods, \textit{e.g.} using dimensional regularization and introducing local counterterms \cite{Peskin:257493}, which maps these divergences onto the RG flow of the Lagrangian parameters. 
The fact that such  divergences do not affect the quantum force  can be shown in complete generality at one-loop using \textit{e.g.} the heat kernel expansion \cite{bordag2009advances}.
Apart from these well-behaved divergences,  other ``spurious'' divergences which are non-removable by renormalization can appear in certain Casimir calculations. This happens for example for a calculation of the pressure in the Dirichlet sphere \cite{Milton_Sphere} (see detailed discussion in Sec.\,\ref{se:Sphere}), and even for 
a version of the calculation of the pressure between plates (see detailed discussion in Sec.\,\ref{se:plane-plane}). 
While such spurious divergences might be easily identified and removed in an ad hoc way,  in this work we will show that they are systematically removed by the requirement of conservation of matter in the sources.  
Matter conservation has so far not been taken into account in calculations of quantum forces,  to the best of our knowledge. At the conceptual level it is the  new ingredient of our  calculations.

Hopefully our presentation will be useful to cosmologists who would like to see their models tested in the laboratory. 
In particular we  transparently explain in terms of Feynman diagrams  how the screening of the scalar inside the sources gives rise to two distinct regimes for the quantum force. This situation is  similar to the Casimir and Casimir-Polder forces  arising in electromagnetism. The common nature and unified description of these electromagnetic quantum forces has been long known, see \textit{e.g.} Refs.\,\cite{Dzyaloshinskii_1961,Schwinger:1977pa}. 
Here we will give a formula  for the quantum force in the massive scalar case and relate the ``scalar Casimir'' and ``scalar Casimir-Polder'' forces.\,\footnote{See \textit{e.g.} \cite{Feinberg:1968zz,FeinbergSucherNeutrinos,Grifols:1996fk,Fichet:2017bng,Costantino:2019ixl} for Casimir-Polder-like forces in the context of particle physics.  }
The use of this general formula is mandatory for computing certain observables such as phase shifts in interferometry.

 Finally, another purpose of this work is to lay out the foundations for more phenomenological studies, for which the effect of the quantum force from a  particle beyond the Standard Model needs to be predicted in realistic experiments. As such, we will present  a calculation of the phase shift induced by a quantum force and measurable in atom interfero\-meters. 
We will then study to which extent atom interferometry is a competitive method to search for light dark particles.

\subsection*{Technical Review}

The present study bears some technical similarities  with other works from the Casimir literature. Here we briefly list a few of these references. 
Our definition of the quantum work involves the variation of the quantum vacuum energy with respect to a deformation of the source. A similar approach relying on energy variation has been used in \cite{Schwinger:1977pa}, expressed in the language of Schwinger's source theory. 
  The source we consider has finite density, and the Dirichlet limit is obtained  by sending the density to infinity.  A similar framework was considered in \cite{Graham:2002fw,Graham:2002xq,Graham:2003ib} (the conservation equation is not exploited in these references).  
In our formalism we consider arbitrary deformations of an arbitrary body. A somewhat similar approach was taken in \cite{Schwinger:1977pa} in the case of deformation of a dielectric fluid. 
We sometimes use regularisation by point-splitting, in the context of Casimir calculations this method  has for example been used in \cite{Milton:2004ya}.
Finally, under certain conditions including that the  bodies are incompressible, finiteness of the quantum work at one-loop can be shown using the heat kernel formalism. This is reviewed and discussed in App.\,\ref{app:HK}.   
\\
\textit{Comparison to \cite{Franchino-Vinas:2021lbl} :} After pre-publication  we became aware of the recent article \cite{Franchino-Vinas:2021lbl} whose scope and results have partial  overlap with the present study --- upon appropriate matching of terms and concepts. 
Both works introduce the concept of variable/effective mass. 
The \textit{principle  of virtual work} whose proof for arbitrary geometry in a specific model is presented in \cite{Franchino-Vinas:2021lbl} is fully compatible with the general formula of the \textit{quantum work} presented here.  
The translation of the key quantities between the model of \cite{Franchino-Vinas:2021lbl} and the present work is $ \phi|_{[\color{blue}{35}]} \equiv \Phi$,
$\frac{\lambda_1}{4}\sigma^2\phi^2|_{[\color{blue}{35}]}\equiv {\cal B}_m J$. 
An important difference  of scope is that in  \cite{Franchino-Vinas:2021lbl} the deformation flow of the body is rigid, which implies that no discussion of matter conservation is needed. In contrast, the compressible deformation flow and its relation to matter conservation are a key aspect of our study. 
Finally, the formula for the quantum work resulting from the rigid deformation of an interface is obtained in \cite{Franchino-Vinas:2021lbl} in terms of a surface integral of the stress tensor. This result  turns out to precisely match the one we obtain in our Eq.\,\eqref{eq:Wshell3} upon setting the variation of density to zero,  even though the intermediate steps of the derivations are rather different. This is a nontrivial verification of the  consistency  of both works.


\subsection*{Outline}

The paper is arranged as follows. 
Section \ref{se:quantumwork} presents the  framework for scalar quantum forces, giving a formula for the quantum work valid at the non-perturbative level for arbitrary deformable sources. 
Section \ref{se:WeakCoupling} specializes  to the weak coupling case (which includes the case of effective field theories). The quantum work in the limiting  case of thin-shell geometries is further evaluated. 
Section \ref{se:Casimir} specializes  to two rigid bodies, showing that the 
Casimir and Casimir-Polder forces for massive scalar fields are asymptotically recovered as limits
of our unifying formula for quantum forces. 
The Casimir pressure on the Dirichlet sphere  is revisited and shown to be finite in section \,\ref{se:Sphere}. The generalized Casimir forces for finite density objects and a finite vacuum mass for the scalar field in plane-plane and point-plane geometry are respectively computed  in sections \ref{se:planar}. In section \ref{se:interferometry} we then  calculate the  phase shift in atom interferometers.
In App.\,\ref{app:CP} we compare a computation of the Casimir-Polder interaction from a scattering amplitude and our derivation, showing explicitly that they coincide. In App.\,\ref{app:HK} we review a proof of the finiteness of the quantum work at the one-loop level {using} the heat kernel expansion following the steps of \cite{bordag2009advances}, which contributes to motivating our approach.   
App.\,\ref{se:loop_divergence} contains details on a loop calculation in momentum space .

\subsection*{Definitions and Conventions}

We assume $d+1$-dimensional Minkowski spacetime ${\cal M}_{d+1}$ with mostly-minus signature $(+,-,\ldots,-)$. The $d+1$ Cartesian coordinates are denoted by $x_\mu$, spatial coordinates are denoted by $x_i\equiv\x$. 
We will be considering a source $J(\x)$ with arbitrary shape and dimension. The support of the source is described by the indicator function ${\bm 1}_J(\x)$ or equivalently using a continuous support function $l(\x)$ which is positive where $J$ is supported  and negative where it is not, with ${\bm 1}_J(x)\equiv \Theta(l(\x))$ where $\Theta$ is the Heaviside distribution. 
The boundary of the source is denoted by $\partial J=\left\{\x\in {\cal M}_{d+1}|l(\x)=0\right\}$. 
Integration over the support of the source $J$ is denoted $\int_J d^d\x$. Integration over the support of the boundary  $\partial J$  is denoted $\int_{\partial J} d\sigma(\x)$. {Although our main focus is ultimately the  $d=3$ case, we  keep $d$ general when possible and specialize to $d=3$ in specific settings.}

\section{The Quantum Work}

\label{se:quantumwork}

In this section we compute the quantum work felt by a source bilinearly coupled to a quantum field under an arbitrary deformation of the source. 
We focus on a scalar field $\Phi$ for simplicity --- generalization to spinning fields is identical although more technical.  
The bodies subject to the Casimir forces are assumed to be classical and static. The set of bodies is collectively represented in the partition function by a static source term $\J(\x)$. More precisely, the $J(\x)$ distribution corresponds to the vacuum expectation of the density operator $\hat n(\x)$ in the presence of matter, 
$J(\x)=\langle\Omega|\hat n(\x)|\Omega\rangle$.

\subsection{Action and Quantum Vacuum Energy } 

We consider the fundamental Lagrangian
\be
{\cal L}[\Phi] = \frac{1}{2}(\partial_M\Phi)^2-\frac{1}{2}m^2 \Phi^2+\ldots \,. 
\ee
The ellipses include possible interactions of $\Phi$, which do not need to be  specified. The interacting theory for $\Phi$ can either be renormalizable --- with either weak or strong coupling, or may also be an effective field theory (EFT) involving a series of operators of arbitrary dimensions.  In this latter case the theory is weakly coupled below the EFT cutoff scale on  distances larger than $\Delta x \sim \frac{1}{\Lambda}$. The scale $\Lambda$ is the energy cut-off of the theory. 

We consider the partition function in Minkowski spacetime\,\footnote{We  call the generating functional $Z[J]$ the partition function in analogy to the Euclidean case.}
\be
Z[J]=\int {\cal D} \Phi e^{ i \left(S[\Phi] - \int d^{d+1}x  \, {\cal B}[\Phi] \J(\x) \right)}
\label{eq:Z_def}
\ee
where 
${\cal B}[\Phi]$ is a bilinear operator in $\Phi$. This operator can encode an arbitrary number of  field derivatives. We distinguish two cases. 
If the bilinear operator has no derivative, then the scalar theory can be renormalizable. 
In this case we write the operator as
\be
{\cal B}_{m}[\Phi]=  \frac{1}{2\Lambda}\Phi^2  \label{eq:Bren}
\ee
If the bilinear operator has  derivatives, then the scalar theory is an EFT and in general contains a whole series of higher dimensional operators. 
In this case, including an arbitrary number of such terms,  we can write the operator as
\be
{\cal B}_{\rm EFT}[\Phi]= \sum_{n>1,i}  \frac{c_{n,i}}{2\Lambda^{2n+1}} \Phi {\cal O}_{n,i} \Phi  \,   \label{eq:BEFT}
\ee
where ${\cal O}_{n,i}$ is a  scalar derivative operator with $2n$ derivatives which  act either to the left or to the right,\,\footnote{When  writing the bilinear interaction in the form Eq.\,\eqref{eq:BEFT},  consistently with the source term defined  in \eqref{eq:Z_def}, we take into account that any derivative acting on $J$ has been removed using integration by part, producing derivatives that act on the fields.
 With this constraint, the most general structure  is the one given in Eq.\,\eqref{eq:OEFT}.  }
\be
 {\cal O}_{n,i} = (\overleftarrow \partial^{2})^p \overleftarrow\partial_{\mu_1}\ldots \overleftarrow\partial_{\mu_{a}}
 (\overrightarrow \partial^{2})^{n-p-a} \overrightarrow\partial_{\mu_1}\ldots \overrightarrow\partial_{\mu_{a}} \label{eq:OEFT}
\ee
The $c_{n,i}$ coefficients are dimensionless. 

Let us comment further about some aspects of EFT.  
Here we distinguished between the ${\cal B}_{m}$ and ${\cal B}_{\rm EFT}$ operators, essentially  because our subsequent analysis of finiteness of the quantum work slightly differs between both cases. In general, a given EFT can feature both the ${\cal B}_{m}$ and ${\cal B}_{\rm EFT} $ terms. In such case the ${\cal B}_{m}$ contribution would tend to dominate, unless it is suppressed by a symmetry (\textit{e.g.} a shift symmetry). Still in the EFT framework, we may also notice that non-derivative higher order operators involving  powers of $\Phi$ and $J$ are in general present. While the higher order operators which are bilinear in $\Phi$ could be simply accounted in the generic source $J$, the broader view is that the EFT validity domain prevents these operators from becoming important, \textit{i.e.} schematically $\partial/\Lambda\ll 1$.   Finally we emphasize that our subsequent analyses involving UV divergences can be done at the level of the EFT, without having to specify the underlying completion, see \textit{e.g.} \cite{Manohar:1996cq,Manohar:2018aog} for more details on divergences and renormalization in EFT.

Since the source is static, the partition function takes the form
\be 
Z[J]=e^{-i E[J]T} \label{eq:EJ_def}
\ee 
where $E[J]$ is referred to as the \textit{quantum vacuum energy} and
$T$ is an arbitrary time interval specified in evaluating the time integrals. This time scale will  drop from the subsequent calculations.\,\footnote{ 
In Minkowski space we have in general 
$Z=\langle 0 | e^{-i {\cal H}T}| 0 \rangle$, with ${\cal H}$ the Hamiltonian of the system. This is why the eigenvalue $E[J]$ is identified as the vacuum energy. 
We also mention that upon Wick rotation to Euclidean space, $E[J]$  corresponds to the free energy of the system. }
In general,  we can set $J$ as an abstract quantity that can be used to generate the correlators of the theory. In our case, since the source couples to ${\cal B}$ (see  Eq.\,\eqref{eq:Z_def}), taking functional  derivatives  of $E[J]$ in $J$ generates the connected correlators of the composite operator  $\cal B$, \textit{i.e.} $\langle {\cal B}(x_1) {\cal B}(x_2)\ldots \rangle$.
In this work,  we consider  that $J$ represents a physical distribution of matter, \textit{i.e.}  
$J(\x)$ is taken to be the expectation value of the density operator $\hat n(\x)$ in the presence of matter, 
$J(\x)=\langle\Omega|\hat n(\x)|\Omega\rangle$.
For concreteness, one can for instance think of a nonrelativistic fermion density, appearing for example via  $\bar\psi \psi = n(\x)$ or $\bar\psi \gamma_\mu \psi =\delta_{\mu 0} n(\x)$ in the relativistic formulation.\,\footnote{
Higher monomials contributions such as $\frac{(\bar\psi\psi)^2}{\Lambda^3}$ are neglected in this example. We mention that in the presence of screening, \textit{i.e.} in the Dirichlet or Casimir regime defined in next sections, the validity of the EFT tends to be improved because the coupling to the source tends to be suppressed. Validity of the EFT in the presence of screening  has been discussed in \cite{Brax:2018grq}. 
}


\subsection{The Source and its  Deformation  }

The source is parametrized by 
\be
J(\x) = n(\x) {\bm 1}_J(\x)
\ee
and corresponds to a particle number distribution of mass dimension $d$. 
The support of this distribution is encoded in ${\bm 1}_J(\x) = \Theta(l(\x))$ where the continuous function $l(\x)$  is positive where $J$ is supported  and negative where it is not. The number density $n(\x)$ is in general an arbitrary distribution  over the support. The integral $N_J=\int d\mu_i J(\mu_i)$ amounts to the total particle number of the source. 

We then introduce a deformation of the source.  We assume that matter is deformable  \textit{i.e.}  both the support and the number density  can vary under the deformation. We will see that such a generalization from rigid to deformable matter  is necessary in order to ensure that the calculation is well-defined and that no infinities show up. 

\begin{figure}
\centering
	\includegraphics[width=0.7\linewidth,trim={0cm 4cm 0cm 5cm},clip]{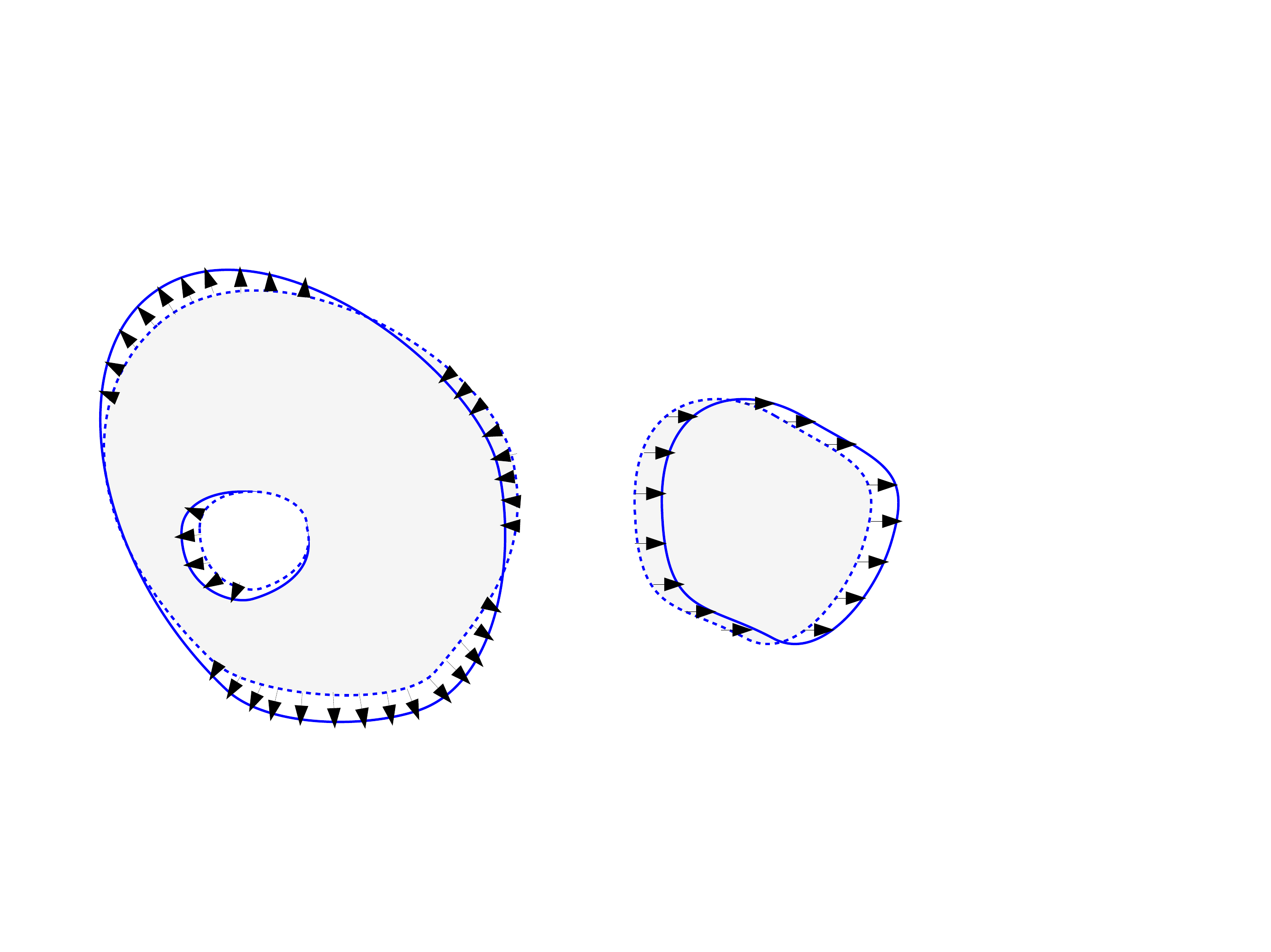}
\caption{ An arbitrary infinitesimal deformation of a generic source. The dotted and plain contours respectively correspond to the boundary of $J_\lambda$ and $J_{\lambda+d\lambda}$. The arrows denote the deformation flow $\Lv$. 
\label{fig:deformation}}
\end{figure}

The infinitesimal deformation of the source is parametrized by a scalar parameter $\lambda$. Under our assumptions the source depends on this deformation parameter as $J_\lambda(\x) = n_\lambda(\x)\Theta[l_\lambda(\x)]$. The deformed source takes the form $J_{\lambda+d\lambda}(\x) = n_{\lambda+d\lambda}(\x)\Theta[l_{\lambda+d\lambda}(\x)]$. 
The deformation of the support of the source is parametrized by 
\be l_{\lambda+d\lambda}(\x)=l_\lambda(\x- \Lv(\x) d\lambda)\ee 
where the $\Lv$ vector is the deformation flow. 
Defining $\frac{\partial}{\partial \lambda}\equiv\partial_\lambda$ the variation of the source under the $\lambda$ deformation is then given by
$\partial_\lambda J_\lambda(\x) =\partial_\lambda n_\lambda \Theta[l_\lambda(\x)] -   n_\lambda\Lv \cdot \bpartial l_\lambda\, \delta[l_\lambda(\x)] $. 
An arbitrary deformation of a generic source is pictured in Fig.\,\ref{fig:deformation}.

 We assume that  $J(\x)$ is  made out of classical matter   and is not a completely abstract distribution.  As the source is made of classical matter, then its \textit{local number density must be conserved}. 
Any deformation of the source must be therefore subject to the conservation of the number density. The local conservation equation under the deformation parametrized by $\lambda$ is
 \be  \partial_\lambda n_\lambda + \bpartial\cdot \left(n_\lambda \Lv \right)= 0 \,.
\label{eq:cons_gen}
\ee
It implies the integral form \be \partial_\lambda \int_{ J_\lambda} d^{d}\x\, n_\lambda(\x) = \int d^{d}\x\, \partial_\lambda  J_\lambda(\x) =  0 \label{eq:cons_gen_int} \ee 
where the second integral is over all space.

{In the special case where the density $n$ is constant in $\lambda$ and $x$, Eq.\,\eqref{eq:cons_gen}  reduces to the condition of an incompressible deformation flow, $\bpartial \cdot \Lv=0$. In this case the source describes  incompressible matter (e.g. a fluid). If $\bpartial \cdot \Lv=0$ and  $\Lv$ is piecewise constant in $x$, then the source describes \textit{rigid} matter. Section.\,\ref{se:Casimir} specializes to such rigid sources. }

\subsection{Quantum Work and Force}

We now study how the quantum system evolves upon a general, infinitesimal deformation of the source, $\lambda\to \lambda+d\lambda$. To proceed we introduce the quantum work under a variation in $\lambda$, 
\be
\W_\lambda = -\partial_\lambda E[J_\lambda]\,. \label{eq:W_def}
\ee
In the particular case of a rigid body and if the deformation field can be factored out, then a quantum force can be defined as 
\be \W_\lambda = \Lv \cdot \F.
\ee
This defines the quantum force between the objects.  Such a simplification is however generally not possible,  the most fundamental quantity to consider is the quantum work. 
Our definition of  quantum work is closely related to variational approaches such as  the principle of virtual work applied to quantum physics, see \textit{e.g.} \cite{Schwinger:1977pa} and \cite{Li:2019ohr,Franchino-Vinas:2020okl} for related recent studies.

The quantum vacuum energy  $E[J]$ is a formally divergent quantity. However, if one varies it with respect to a physical parameter, the resulting variation is a physical observable and thus should be  finite.\,\footnote{Throughout this work, the physical parameter  is typically a geometric variable such as the distance between two bodies. }
Hence even though $\W_\lambda$ encodes all the quantum effects felt by the source(s), the only divergences remaining in this quantity are those with a physical meaning, \textit{i.e.} the ones which must  be treated in the framework of renormalization. 
One origin for such divergences is  the field interactions, another can be the curvature of spacetime, as pointed out in \cite{Fichet:2021xfn} in AdS. Even for a free theory in flat space, there can  be renormalization of surface tension terms, as pointed out in \cite{Bordag:2004rx}.  
We refer to such divergences as ``physical'' ones --- these are the familiar divergences from QFT, that only appear for specific integer values of spacetime dimension. 
One of our goals in the following is to show that there are no other divergences than the physical ones in the QFT calculation of the quantum force.

Using the definition of Eqs.\,\eqref{eq:Z_def}, \eqref{eq:EJ_def}, the quantum work is given by 
\be
W_\lambda = 
  - \int d^{d} \x \left\langle {\cal B} \right\rangle_{J_\lambda}(x,x)\partial_\lambda J_\lambda(\x) \label{eq:Wdefgen} \,
  \ee
  where $\left\langle {\cal B}\right\rangle_{J_\lambda}$ is the time-ordered quantum average of ${\cal B}$ in the presence of the source $J_\lambda$. 
Here and throughout we use the shortcut  notation $W_{J_\lambda}=W_\lambda$, 
  $\left\langle {\cal B}\right\rangle_{J_\lambda}=\left\langle {\cal B}\right\rangle_{\lambda}$.  While all quantities depend on $\lambda$, the $\lambda$ dependence is relevant only under the $\partial_\lambda$ variation and is often dropped elsewhere. The two-point function in presence of the source at coinciding points will be sometimes denoted by $\left\langle {\cal B}\right\rangle_{J}$ in the rest of the paper. 
  
Using the general form of the bilinear coupling, ${\cal B}$ is expressed in terms of the two point function of $\Phi$ in the presence of $J$, $\langle T \Phi(x)\Phi(y)\rangle_J$, evaluated at coinciding points.  
 We assume that the classical value of $\Phi$ is zero.\,\footnote{This can be the consequence of a $Z_2$ symmetry, enforcing that $\Phi$  only appears bilinearly in the action such that $\langle\Phi\rangle=0$. In the presence of nonzero $\langle\Phi\rangle$, the possible $Z_2$ symmetry is broken, implying that at weak coupling the fluctuation of $\Phi$ over the $\langle\Phi\rangle$ background has a linear coupling to the source. As a result a classical force is also present in addition to the quantum force. This case does not need to be investigated in the present paper.  This extra classical force appearing in the presence of $\langle\Phi\rangle\neq 0$ does not automatically dominate over the quantum force --- instead, various regimes arise  as a function of the geometry.   These aspects have been partly investigated  in \cite{Brax:2018grq}.  } Hence the disconnected part of the two-point function vanishes and the two-point function  reduces to $\langle T \Phi(x)\Phi(y)\rangle_J = \Delta_J(x,y)$, where $\Delta_J(x,y)$ is the Feynman propagator in the presence of the source $J$.
 The quantum average of ${\cal B}$ is then expressed in terms of the Feynmann propagator $\Delta_J$  as 
\be
\langle {\cal B}_{ m }\rangle_J=  \frac{1}{2\Lambda}\Delta_J(x_1,x_2)  \label{eq:Bvevren}
\ee
\be
\langle{\cal B}_{\rm EFT}\rangle_J= \sum_{n>1,i}  \frac{c_{n,i}}{2\Lambda^{2n+1}}  {\cal O}_{n,i} \Delta_J(x_1,x_2) \,.   \label{eq:BvevEFT}
\ee
Here in the EFT case we introduce the shortcut  $\langle T \Phi(x) {\cal O}_{n,i} \Phi(y)\rangle_J \equiv {\cal O}_{n,i} \Delta_J(x,y)$ where the ${\cal O}_{n,i}$ operator as defined in Eq.\,\eqref{eq:OEFT} has derivatives acting to the left and to the right. 
The general formula for the quantum work presented in Eq.\,\eqref{eq:Wdefgen} is valid at the non-perturbative level.

\subsection{ $\W_\lambda$ is Finite (Renormalizable Case)}

\label{se:W_finite}

The quantity $\left\langle {\cal B}\right\rangle_J(x,x)$ is formally divergent since it contains the propagator at coinciding points. It is thus not obvious why the quantum work $\W_\lambda$ should be finite. 
Throughout this subsection, we  regulate the divergence in $\left\langle {\cal B}\right\rangle_J$ by introducing a small splitting of the endpoints, $\left\langle {\cal B}\right\rangle_J(x,x)\equiv\left\langle {\cal B}\right\rangle_J(x,\xeps)|_{\epsilon\to 0}$ where we defined $\xeps=x+\epsilon$.\,\footnote{  
The analogous regularization in Fourier space is  momentum cutoff,  $p<\frac{1}{\epsilon}$. These regularizations admit a physical meaning.  $\epsilon$ can be thought as the distance scale below which the description of the classical matter as a continuous distribution breaks down. In this view, the cutoff length $\epsilon$ has  a physically meaningful value, and the statement of  existence of a divergence  turns into a statement on  $\epsilon$-dependence of the result. }

We consider the renormalizable case \textit{i.e.} the $\langle{\cal B}\rangle_m$ operator (the EFT case is addressed in section \ref{se:Wfinite_EFT}).  
We assume that, as a preliminary step, all the divergences which can be removed by the renormalization of the coupling constants of local operators have been performed, \textit{e.g.} the fundamental mass  $m$ is the renormalized mass.
We assume that $n(\x)$ is \textit{finite} for any $\x$ over the support of $J$ (the $n\to \infty$ limit is discussed in section \ref{se:Dirichlet}).  No assumption on the interactions of $\Phi$ is necessary thus $\Phi$ can be strongly coupled. 
{Under these conditions the  finiteness of the quantum work can be shown as follows.}

In the $\epsilon\to 0$ limit, we can decompose the expectation value of $\cal B$ as the sum of \textit{i)} the $\epsilon$-dependent, would-be divergent term, and  \textit{ii)} a  finite term in which the $\epsilon$ dependence amounts to $O(\epsilon)$ corrections that can be neglected for $\epsilon\to 0$. This gives
\be
\left\langle {\cal B}\right\rangle_J(x,\xeps) = \left\langle {\cal B}\right\rangle_J^{\rm div }(x,\xeps)+ \left\langle {\cal B}\right\rangle^{\rm fin}_J(x,x) + O(\epsilon)  \,.  \label{eq:Bren_dec}
\ee 
We then use the assumption that the number density is finite. It implies that the effective squared mass of $\Phi$ inside the source, $m^2+\frac{n(\x)}{\Lambda}$, is finite. We remind that $m$ has been renormalized already.  
A divergence {{in the short distance behaviour of $\langle{\cal B}\rangle(x,\xeps)$ }} can arise from  a  propagator going from $x$ to $\xeps$. As the effective mass term amounts to a \textit{relevant} operator with finite value, it is negligible in the short  distance limit of the propagator, \textit{i.e.} in the large momentum limit.
As the source $J$ appears in  the propagator only via the effective mass term, we conclude that  the  divergent piece  $\left\langle {\cal B}\right\rangle_J^{\rm div }(x,\xeps)$ is independent of $J$. Furthermore, in that short distance limit, the propagator must be Lorentz invariant, and  we conclude that the divergent piece in Eq.\,\eqref{eq:Bren_dec} is independent of $x$, 
\be 
\left\langle {\cal B}\right\rangle_J^{\rm div }(x,\xeps)|_{{\rm small}\,\epsilon } =\left\langle {\cal B}\right\rangle^{\rm div }(x,\xeps) =\left\langle {\cal B}\right\rangle^{\rm div }(0,\epsilon)\equiv  \left\langle {\cal B}\right\rangle^{\rm div }_\epsilon \,. \label{eqBeps}
\ee
We thus see that the divergent piece  depends only on $\epsilon$ and diverges in the $\epsilon \to 0 $ limit. 
Using the decomposition Eq.\,\eqref{eq:Bren_dec}  and  the definition of the quantum work, we obtain the decomposition 
\be 
W_\lambda = W^{\rm fin}_\lambda + W^{\rm div}_\lambda 
\ee
with
\be
W^{\rm fin,div}_\lambda = 
  - \int d^{d} \x \left\langle {\cal B}\right\rangle_{J}^{\rm fin,div}(x,x)\partial_\lambda J_\lambda(\x) \,. 
\ee
In the divergent piece, $\left\langle {\cal B}\right\rangle^{\rm div }_\epsilon$ factors out of the integral because it is independent of $x$. This gives
\be
W^{\rm div}_\lambda = 
  - \left\langle{\cal B} \right\rangle_\epsilon^{\rm div }\int d^{d} \x \, \partial_\lambda J_\lambda(\x) \label{eq:Wdiv} \,.
  \ee
The remaining integral corresponds exactly to the variation of the total density of the source under the deformation, appearing in the integral form  of the conservation equation Eq.\,\eqref{eq:cons_gen_int}. Thus if the equation of conservation Eq.\,\eqref{eq:cons_gen_int} is satisfied, then Eq.\,\eqref{eq:Wdiv} vanishes. 
We conclude that, upon conservation of matter in the source, for any deformation and finite $n$ the quantum work is finite: 
\be
W^{\rm div}_\lambda = 0\,. 
\label{eq:Finiteness}
\ee
This is true at the nonperturbative level. 
In the particular case of incompressible matter, Eq.\,\eqref{eq:Wdiv} reduces to 
\be
W^{\rm div}_\lambda|_{\rm incompressible} = 
  - n \left\langle{\cal B} \right\rangle^{\rm div }\int_{J} d^{d} \x \, \bpartial \cdot L(\x) \label{eq:Fdivrigid} \,.
  \ee
In that case we can say that, upon conservation of matter in the source, for any divergent-free deformation flow  and finite $n$, the quantum work is finite: 
\be
\W^{\rm div}_\lambda|_{\rm incompressible} = 0 
\label{eq:Finiteness2}
\ee
The finiteness of the quantum work will be exemplified in the upcoming sections. 

Finally we comment on the finite part of the quantum work.  The finite part can be put in the useful alternative form by evaluating the integrand, using the divergence theorem and using the conservation equation,
\be
W^{\rm fin}_\lambda = 
   - \int d^{d} \x  \,n_\lambda(\x) \Lv \cdot \bpartial \left[ \left\langle {\cal B} \right\rangle_\lambda(x,x) \right] \label{eq:Wfinren} \,.
\ee
This is another way to verify that any constant piece in $\left\langle{\cal B} \right\rangle(x,x)$ does not contribute to the quantum work as it appears under a gradient  in the integrand.

\section{Weak Coupling: Finiteness Properties and Thin Shell Limit}
\label{se:WeakCoupling}

At weak coupling the $\Phi$ field has an equation of motion (EOM) that we can use to  evaluate the quantum work. 
We introduce the bilinear operator ${\cal B}''$, defined by
\be
{\cal B} = \frac{1}{2} \Phi(x){\cal B}'' \Phi(x)\,. 
\ee
This is the operator that appears in the EOM. 
For example, when applied to ${\cal B}_m$ this is ${\cal B}''_m=\frac{1}{\Lambda} $. 
In the EFT case, ${\cal B}''$ is the differential operator appearing in Eq.\,\eqref{eq:BEFT}, 
\be
{\cal B}_{\rm EFT}''= \sum_{n>1,i}  \frac{c_{n,i}}{\Lambda^{2n+1}} {\cal O}_{n,i} \,. 
\ee
The left and right derivatives in ${\cal O}_{n,i}$ act on the propagators attached respectively to the left and right of the vertex. 

At leading order in the perturbative expansion, the $\Delta_J(x,x')$ propagator
satisfies the equation of motion  \be {\cal D}_x \Delta_J(x,x') + {\cal B}'' J(\x) \Delta_J(x,x') = -i \delta^{d+1}(x-x') \label{eq:EOMgen} \ee where ${\cal D}=\square+m^2$ is the wave operator and $\square$ is the scalar d'Alembertian.
The solution to Eq.\,\eqref{eq:EOMgen} is a Born series that describes  the bare propagator $\Delta_0$ (\textit{i.e.} $\Delta_J|_{J\to 0}$)   dressed by insertions of ${\cal B}''J$. For convenience we define  the insertion \be \Sigma(x,y)=-i{\cal B}''J(\x) \delta^{d+1}(x-y)\ee
and we use the inner product $f\star g = \int d^{d+1}u\, f(u)g(u)$. With these definitions the dressed propagator is given by 
\begin{align}
\Delta_J(x,x') &= \sum^\infty_{q=0} \Delta_0 \left[\star \Sigma \star \Delta_0\right]^q(x,x')
\\
&=
\Delta_0(x,x') -  \int d^{d+1} u \,\Delta_0(x,u) i {\cal B}'' J({\bm u})  \Delta_0(u,x')+\ldots
\end{align}

\begin{figure}
\centering
	\includegraphics[width=0.9\linewidth,trim={0cm 5cm 0cm 6cm},clip]{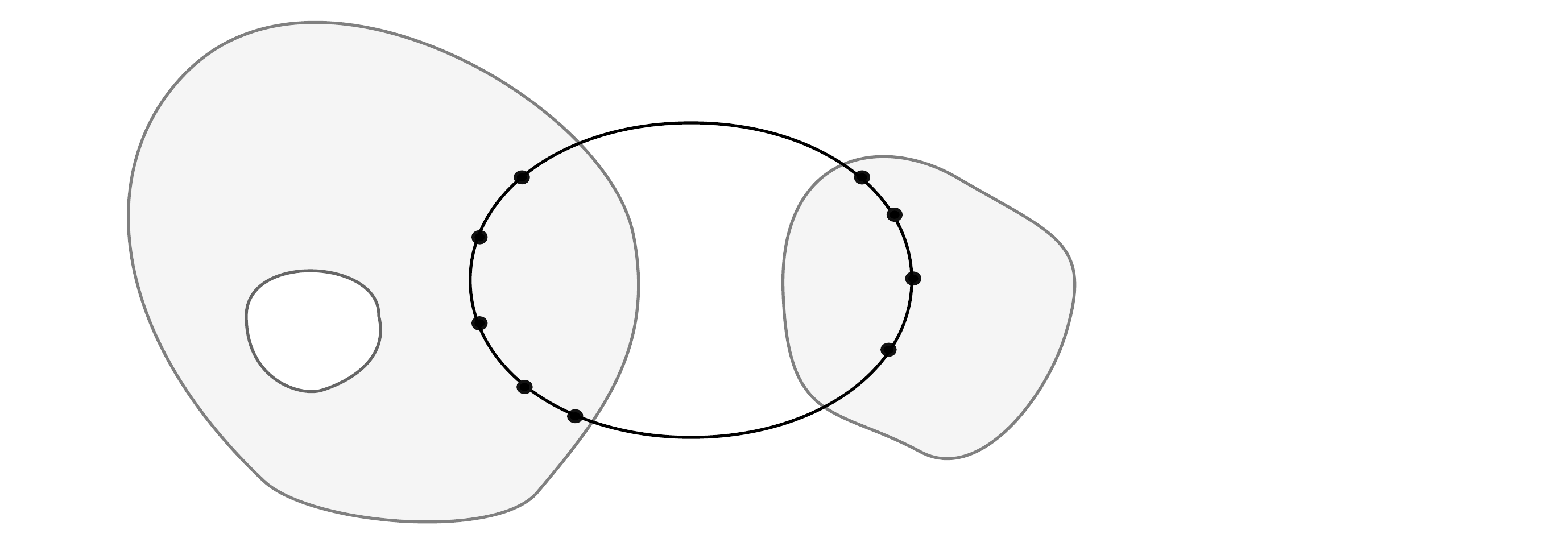}
\caption{ A sample  one-loop  diagram in the presence of an arbitrary source.
Black dots represent insertions of $-i{\cal B}'' J(\x)$. Each black dot is integrated over the support of the source. Under an infinitesimal deformation of the source ($\lambda\to\lambda+d\lambda$), the corresponding variation (\textit{i.e.} $\partial_\lambda$) of this diagram contributes to the one-loop quantum work Eq.\,\eqref{eq:F1loop1}. 
\label{fig:diagrams_1loop}}
\end{figure}

Putting  this result back into Eq.\,\eqref{eq:Wdefgen} provides the leading, one-loop contribution to the quantum work,\,\footnote{
Another, conceptually similar way to derive this formula is via the heat kernel formalism, see \textit{e.g.} \cite{Bordag:2004rx,Franchino-Vinas:2020okl}
}
\be
\W_\lambda^{\rm 1-loop} = -\frac{1}{2\Lambda} \int d^{d}{\x} {\cal B}'' \sum^\infty_{q=0}  \Delta_0 \left[\star \Sigma \star \Delta_0\right]^q(x,x) \partial_\lambda J(\x)
\label{eq:F1loop1}
\ee
This is valid for both  ${\cal B}_m$ and ${\cal B}_{\rm EFT}$ insertions. 
In terms  of Feynman diagrams Eq.\,\eqref{eq:F1loop1} is simply a loop with an  arbitrary number of insertions of ${\cal B}''J$ and one insertion of $\partial_\lambda J$.  A term of the series is represented (without the  $\partial_\lambda$ variation) in Fig.\,\ref{fig:diagrams_1loop}, where each insertion is represented by a black dot.

In the EFT case, the validity domain of the EFT implies that higher derivatives terms in ${\cal B}_{\rm EFT}$  must remain small. Still, in the series of Eq.\,\eqref{eq:F1loop1}, the effect of lower derivative terms may become important in some regime. When the effect of these derivatives is important, it may happen that  convergences issues in  Eq.\,\eqref{eq:F1loop1} appear, that need careful consideration taking into account EFT validity. A related example is discussed in details in \cite{Bordag:2001ta}.   In the following, we will show the finiteness of every term in $\W_\lambda^{\rm 1-loop}$ and simply assume that the overall series is convergent.

\subsection{$W_\lambda$ is Finite  (EFT Case)}

\label{se:Wfinite_EFT}

Our proof of Eq.\,\eqref{eq:Finiteness} uses that the effective mass is a relevant operator that becomes negligible at short distances. 
In contrast, the insertions from ${\cal B}^{\rm EFT}$ correspond to irrelevant operators hence the same reasoning cannot apply --- the operators become more important at short distance. 
The solution to this apparent puzzle is that the EFT in its domain of validity is necessarily weakly coupled, hence instead of using a non-perturbative argument one can use the series representation Eq.\,\eqref{eq:F1loop1} to prove finiteness. 

Let us verify finiteness term-by-term. We single out a term from Eq.\,\eqref{eq:F1loop1} and introduce point-splitting, ${\cal B}''  \Delta_0 \left[\star \Sigma \star \Delta_0\right]^q(x,\xeps)$, with $\xeps=x+\epsilon$. Our goal is to show that the divergent piece in this quantity is independent of $x$. 
The term is
\be
{\cal B}''  \Delta_0 \left[\star \Sigma \star \Delta_0\right]^q(x,\xeps) = (-i)^q \left(\prod^{q}_{i=1} \int d^d\bmu_i J(\bmu_i) \int dt_i \right)  \prod^{q}_{i=0} {\cal B}''\Delta_0(\mu_i,\mu_{i+1})\bigg|_{\mu_0=x, \mu_{q+1}=\xeps} \label{eq:loop_points}
\ee
It is understood that one of the block of derivatives in ${\cal B}''$ acts to the left and the other acts to the right. 

The divergence in the diagram defined by Eq.\,\eqref{eq:loop_points} occurs when all  the positions coincide.
This is more easily verified in momentum space, in which case there is  a single loop integral over the internal momentum flowing  around the  loop.  The divergence is tied   to the  large momentum for all the propagators,  which  in position space  corresponds to  the limit of coincident endpoints. 
We present the explicit momentum space calculation   in App.\,\ref{se:loop_divergence}. The divergent piece of Eq.\,\eqref{eq:loop_points} is $(-i)^q c_J^q  L^{\rm div}_{q,\epsilon} $ where $c_J$ is finite and $L^{\rm div}_{q,\epsilon}$ is the divergent part. The key point is that $L^{\rm div}_{q,\epsilon}$  is position-independent.

Putting   this piece into the definition of the quantum work 
Eq.\,\eqref{eq:F1loop1} gives the divergent piece of the quantum work, 
\be
\W_\lambda^{\rm 1-loop, div} = -\frac{1}{2\Lambda} \sum_{q=0}^\infty (-i c_J)^{q} L^{\rm div}_{q,\epsilon} \int d^{d}\x  \partial_\lambda J \,.
\label{eq:F1loop1div}
\ee
The remaining integral corresponds exactly to the variation of the total density of the source under the deformation, appearing in the integral form  of the conservation equation Eq.\,\eqref{eq:cons_gen_int}. Thus if the equation of conservation Eq.\,\eqref{eq:cons_gen_int} is satisfied, then Eq.\,\eqref{eq:F1loop1div} vanishes. 

It follows that, upon conservation of matter in the source, for any deformation and finite $n$ the quantum work is finite: 
\be
\W_\lambda^{\rm 1-loop, div} = 0\, . 
\label{eq:Finiteness1loop}
\ee
The {incompressible} version of this finiteness property trivially follows, like for  Eq.\,\eqref{eq:Finiteness2}.

\subsection{The Thin Shell Limit }

\label{se:Dirichlet}

So far we have considered a generic source as an arbitrary volume in $d$-dimensional space. 
Here we investigate a subset of sources  for which the support is a thin shell approaching  a codimension-one hypersurface.  

We denote the source by $J_{\eta} = n(x) {\bm 1}_{\S_\eta,\lambda}(x) $ where $\eta$  parametrizes the small width of the shell. For $\eta\to 0$ the support of the shell  tends to a hypersurface denoted by $\S $. To avoid any ambiguity we always keep $\eta$ small but nonzero in the following calculations. 
The volume element can be split as \be \int_{\S_\eta}d^d\x \overset{{\rm small}\,\eta}{=} \int_{\S}d\sigma(\x)\int_{\rm width} dx_\perp
\label{eq:VolShell}
\ee 
where the $x_\perp$ coordinate parametrizes the direction normal to $\S$. The boundary of ${\cal S}_\eta$ can also be decomposed as  \be \partial {\cal S}_{\eta\to 0} ={\cal S}_{\rm in}\cup {\cal S}_{\rm out}
\label{eq:partialS}
\ee
where ${\cal S}_{\rm in,out}$ are the two hypersurfaces bounding the volume  enclosed by $\cal S_\eta$ in the limit $\eta \to 0$. The propagator in the presence of the thin shell is denoted by $\Delta_{\cal S}(x,x')$. 
The density can be chosen to scale with $\eta$ such that it remains finite for $\eta\to0$. This happens if the density scales as  $\eta n =$cst.  
In the following, the deformation of the source is kept arbitrary. All quantities depend on the deformation parameter $\lambda$. We will drop the $\lambda$ index when appropriate. 

We evaluate the quantum work for this specific class of sources, taking $\eta$ small but finite. 
For simplicity we consider the coupling to the source induced via the ${\cal B}_m$ operator. 
Our evaluation involves various manipulations of the EOMs and  of the divergence theorem. We emphasize that we do \textit{not} use   the conservation equation. That way we can  explicitly demonstrate later on, at the level of   applications, that the conservation equation is required to obtain finiteness of the quantum work. For clarity we split the calculation in various steps. 

\paragraph{Step \textit{1:}  }

Starting from the general expression of the quantum work Eq.\,\eqref{eq:Wdefgen}, we evaluate the $\partial_\lambda J_\lambda$ variation. We use $\frac{\bpartial l}{\lVert\bpartial l \lVert} = \n_{\rm in} $ with $ \n_{\rm in}$ the inward-pointing normal vector, then use the divergence theorem $\int_{\partial {\cal S}} d\sigma(\x) \n_{\rm out} \cdot {\bm f}(\x) = \int_{ {\cal S}} d^d\x \bpartial\left( {\bm f}(\x) \right) $ with $ \n_{\rm out}= -\n_{\rm in}$. We obtain
\be
W_{{\cal S}_\eta,\lambda} = -\frac{1}{2\Lambda} \int_{\S_\eta} d^d \x\, \Delta_{\cal S}(\x,\x)\partial_\lambda n_\lambda(x)
- \frac{1}{2\Lambda}\int_{\S_\eta} d^d \x \,\bpartial\left[ \Lv n_\lambda(x) \Delta_{\cal S}(x,x)
\right] \,. \label{eq:Wshell1}
\ee

\paragraph{Step \textit{2}: (Using the equations of motion) }

We will further simplify the second term of Eq.\eqref{eq:Wshell1} by observing that it can be related to  discontinuities determined from the equation of motion.  
  In order to proceed we introduce the notation for derivatives acting on either the first or second argument of the propagator, $\bpartial_1 \Delta(x,x')\equiv \bpartial_{x} \Delta(x,x')$, $\bpartial_2 \Delta(x,x')\equiv \bpartial_{x'} \Delta(x,x')$.  The coincident-point propagator is regularized via point-splitting using $\xeps=x+\epsilon$ where the shifted point is taken to belong to ${\cal S}$ when $x\in {\cal S}$. 

In the thin shell limit,  the second term in Eq.\,\eqref{eq:Wshell1}   takes the form 
\be
- \frac{1}{2\Lambda} \eta\,\int_{\cal S} d\sigma(\x)\left(   \bpartial_1[\Lv(x) n(x) \Delta_{\cal S}(x,x_{\epsilon})])+ \Lv(x) n(x) \bpartial_2[  \Delta_{\cal S}(x,x_{\epsilon})]\right)(1 +O(\eta)) 
\label{eq:2ndterm}
\ee
We used the volume element Eq.\,\eqref{eq:VolShell} and that the integrand is continuous over $\S_\eta$.  
In the first term of Eq.\,\eqref{eq:2ndterm} the vector $\Lv$ is kept inside the derivative for further convenience. 

The EOM is
\be {\cal D}_x \Delta_{\cal S}(x,x') + \frac{1}{\Lambda} J_\eta(\x) \Delta_{\cal S}(x,x') = -i \delta^{d+1}(x-x') \label{eq:EOMshell}
\ee
where $\D_x=\square_x+m^2 $.  Below the $\delta^{d+1}$ is always zero due to point splitting. Each of the terms in Eq.\,\eqref{eq:2ndterm} can be expressed using an appropriate derivative of the EOM, with the remaining endpoint set to an appropriate value. 

The first term  in Eq.\,\eqref{eq:2ndterm}  is obtained by multiplying the EOM with $\Lv(\x)$ then applying $\bpartial_x$. We then integrate across the normal coordinate and set the remaining endpoint of the propagator $x'$ to coincide with  $x$ in the transverse coordinates. This gives  the identity
\be
- \frac{\eta}{\Lambda}   \bpartial_1 \left[ \Lv  n  \Delta_{\cal S}(x,x_{\epsilon})\right] (1+O(\eta))
\overset{{\rm small}\,\eta}{=}
\left[\int_{\rm width} dx_\perp    \bpartial_x\left[\Lv(\x)\square_x\Delta_{\cal S}(x,x')\right] \right]_{x'\to x_{\epsilon}}   \label{eq:del1box1} 
\ee
The fundamental mass contributes as $O(\eta)$,  it is thus neglected . 
After integrating over the transverse coordinates (\textit{i.e.} applying $\int_{\cal S} d\sigma(\x)$), the l.h.s of Eq.\,\eqref{eq:del1box1} coincides with  the first term of Eq.\,\eqref{eq:2ndterm}. 

The second term in Eq.\,\eqref{eq:2ndterm} is obtained by applying $\bpartial_{x'}$ to the EOM then contracting with $\Lv(\x')$. The subsequent  steps are the same as above and the result is 
\be
 - \frac{\eta}{\Lambda} \Lv  n  \bpartial_2 \left[   \Delta_{\cal S}(x,x_{\epsilon}) \right](1+O(\eta)) 
\overset{{\rm small}\,\eta}{=}
\left[\int_{\rm width} dx_\perp \Lv(\x')  \bpartial_{x'}\left[\square_x\Delta_{\cal S}(x,x')\right]\right]_{x'\to x_{\epsilon}} \label{eq:del2box1} \,. 
\ee
Using the identities \eqref{eq:del1box1} and \eqref{eq:del2box1} in Eq.\,\eqref{eq:2ndterm} one can  eliminate the presence of the density in favor of d'Alembertians.

\paragraph{Step \textit{3}:}

Finally we use the divergence theorem on the right hand side  of both Eqs.\,\eqref{eq:del1box1} and \eqref{eq:del2box1}. In Eq.\eqref{eq:del2box1}  the divergence theorem turns the ${\square}_x$ into $\n \cdot\bpartial$.\,\footnote{Notice that
$x'$ is set to $x_\epsilon$ only outside of the integral. Hence 
the $x'$ dependence of the integrand is irrelevant when applying  the divergence theorem.}
Replacing these results  in  Eq.\,\eqref{eq:Wshell1} we obtain the final form for the quantum work on a thin shell, 
\begin{align}
W_\lambda^{\cal S} = & \nn
- \frac{1}{2}\int_{\S_{\rm in}\cup \S_{\rm out}} d\sigma(\x) \left(
\,n_i L_j \, \partial^i_1 \partial^j_2  \Delta_{\cal S}(x,x_{\epsilon})
+ \n.\Lv\, \square_1 \Delta_\S(x,x_{\epsilon})  
\right) \\ &
- \frac{1}{2\Lambda} \int_{\S_\eta} d^d \x\, \Delta_{\cal S}(x,\xeps)\partial_\lambda n_\lambda(\x) +O(\eta)  \label{eq:Wshell2}
\end{align}
The volume term in the second line encodes the variation of the number density of the source under the deformation. 
Notice that we have not used the conservation equation yet in deriving Eq.\,\eqref{eq:Wshell2}. 
The conservation equation must ensure  that the quantum work is finite, along the line of \eqref{eq:Finiteness}. This will be exemplified in Sec.\,\ref{se:Sphere}.

\subsection{The Dirichlet Limit}

We have so far assumed that $\eta n\equiv n_{\cal S}$ is finite.  $n_{\cal S}$ can be understood as the number density on the hypersurface $\cal S$. In this subsection we take the limit $n_{\cal S}\to \infty$. In this limit we obtain that the propagator \textit{vanishes} anywhere inside ${\cal S}_\eta$, including on the boundaries ${\cal S}_{\rm in}$, ${\cal S}_{\rm out}$, \textit{i.e.} we have $\Delta_\S(x_,x')=0$ for any $x'$ or $x\in {\cal S}$. On the other hand
the derivatives normal to the boundary do  not vanish on the boundary, \textit{i.e.} $\partial_1^\perp \Delta_{\cal S}(x,x')|_{x\in {\cal S}_{\rm in, out}}\neq 0$. \footnote{
These properties are shown at the level of the EOM using field continuity. We integrate the EOM  on any domain crossing ${\cal S_\eta}$ and use the divergence theorem. This makes appear  the well-known "jump"  in the normal derivatives
\be
\partial_1^\perp \Delta_{\cal S}(x,x')|_{x\in {\cal S}_{\rm  out}}-\partial_1^\perp \Delta_{\cal S}(x,x')|_{x\in {\cal S}_{\rm  in}} = 
\frac{n_{\cal S}}{\Lambda}\Delta_{\cal S}(x,x')|_{x\in {\cal S}}\times (1+O(\eta))\,. 
\ee
The propagator is continuous everywhere, hence  on the rhs is it enough to simply write $x\in {\cal S}$ without further detail. 
Due to the requirement of continuity of the propagator, the discontinuity of derivatives must remain finite. As a result, taking the limit $n_{\cal S}\to \infty$ implies that $\Delta_{\cal S}(x,x')|_{x\in {\cal S}}\to 0$ for any $x'$.  }
In summary, the limit $n_{\cal S}\to \infty$ amounts to a \textit{Dirichlet boundary condition} for the propagator. 

We now apply this Dirichlet limit to the quantum work Eq.\,\eqref{eq:Wshell2}. 
The second surface term in the first line of  Eq.\,\eqref{eq:Wshell2} vanishes in the Dirichlet limit since the second endpoint belongs to the boundary and has no derivative acting on it. In contrast the first surface term involves derivatives on both endpoints and thus does not vanish in the Dirichlet limit. 
This term further simplifies: the first derivatives of $\Delta(x,x')$ across ${\cal S}$ are discontinuous only in the normal coordinate, while they are continuous in the other directions, therefore $n_i L_j \, \partial^i_1 \partial^j_2  \Delta_{\cal S}(x,x_{\epsilon}) = L_\perp \, \partial^\perp_1 \partial^\perp_2  \Delta_{\cal S}(x,x_{\epsilon})$. We see that only the normal component of the deformation flow $L_\perp = \n\cdot \Lv $ contributes in the Dirichlet limit.

We can then make contact with the scalar stress-energy tensor $T^{\mu\nu}= -\partial^\mu\phi\partial^\nu \phi+\frac{1}{2}\eta^{\mu\nu}(\partial_\rho\phi \partial^\rho \phi-m^2\phi^2) $. Namely, when considering the normal component $T^{\perp\perp}$, we recognize that the difference of the  time-ordered expectation value $\langle T^{\perp\perp}\rangle$ between ${\cal S}_{\rm out}$ and ${\cal S}_{\rm in}$ is \be
\left[\langle  T^{\perp\perp}   \rangle \right]^{{\cal S}_{\rm out}}_{{\cal S}_{\rm in}}= \frac{1}{2}
\left[\langle  \partial^\perp\phi\partial^\perp \phi  \rangle \right]^{{\cal S}_{\rm out}}_{{\cal S}_{\rm in}} 
=
\frac{1}{2} \left[ \partial_1^\perp \partial_2^\perp \Delta_{\cal S}(x,x) 
\right]^{{\cal S}_{\rm out}}_{{\cal S}_{\rm in}} 
\ee
Notice that there is a contribution from the normal derivatives inside the isotropic $\eta_{\mu\nu}$ term. The derivatives in the transverse directions do not contribute  due to continuity. It follows that the quantum work on a thin shell in the Dirichlet limit can be expressed using the stress-energy tensor as 
\be
W_\lambda^{\cal S} = 
- \int_{\S_{\rm in}\cup \S_{\rm out}} d\sigma(\x) 
\, L_\perp \,  \langle \Omega \vert T^{\perp\perp}\vert \Omega \rangle 
- \frac{1}{2\Lambda} \int_{\S_\eta} d^d \x\, \Delta_{\cal S}(x,\xeps)\partial_\lambda n_\lambda(\x) +O(\eta)  \label{eq:Wshell3}
\ee
This is, of course, not a coincidence.   This contribution to the quantum work reproduces  exactly the difference of stress-energy tensors that is used to compute the  Casimir forces or pressures on  thin shells, see \textit{e.g.} \cite{Milton_Sphere}. 
The second term, which is the new term  arising in our calculation, must ensure that any spurious divergence arising from the first term cancels out upon requiring matter conservation  of the source, as dictated by the finiteness property Eq.\,\eqref{eq:Finiteness}. 
We will exemplify this in the case of the Dirichlet sphere in Sec.\,\ref{se:Sphere}.

\section{Scalar Quantum Forces between Rigid Bodies}

\label{se:Casimir}

\begin{figure}
\centering
	\includegraphics[width=0.9\linewidth,trim={0cm 2cm 0cm 2cm},clip]{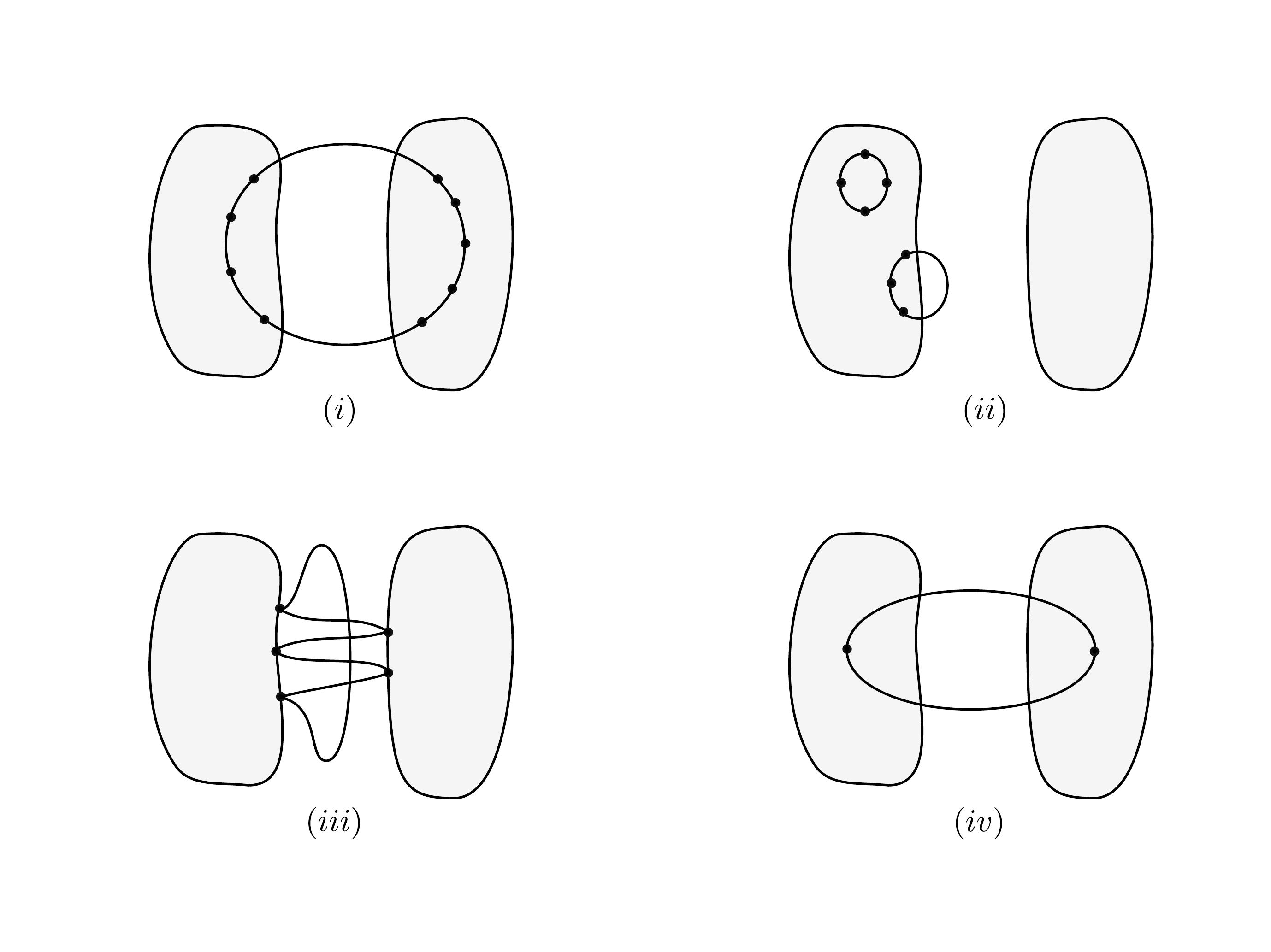}
\caption{ Sample Feynman diagrams for a scalar field in the presence of two extended sources. \textit{(i)}: A generic one-loop contribution. \textit{(ii)}:  ``Tadpole'' loops vanishing under $\partial_\lambda$. \textit{(iii)}: Casimir limit (Strong coupling to sources) \textit{(iv)}: Casimir-Polder limit (Weak coupling to sources)
\label{fig:diagrams}}
\end{figure}

\subsection{Rigid Bodies}

In this section we choose a  specific shape for the source and the deformation. The source is assumed to be the compound of two rigid bodies  $J=J_1+J_2$  with number densities $n_{1,2}$. We assume that the $J_2$ source moves rigidly with respect to $J_1$. The deformation flow $\Lv$ thus  reduces to a constant vector  over $J_2$ and vanishes elsewhere. In this particular case the $\Lv$ factors out in $\W_\lambda$ and we can talk about the quantum force ${\bf F}_{1\to 2}$ between $J_1$ and $J_2$. 
Using Eq.\,\eqref{eq:F1loop1} the general expression for the quantum work is expressed as
\be
\W_\lambda^{\rm 1-loop} = \Lv \cdot {\bf F}_{1\to 2} = -  \frac{1}{2\Lambda} \int d^{d}\x \sum^\infty_{q=0}  \Delta_0 \left[\star \Sigma \star \Delta_0\right]^q(x,x) \Lv \cdot \bpartial J_2(x)
\label{eq:F1loopCas}
\ee
where $\Sigma = -\frac{i}{\Lambda}(J_1+J_2)\delta^{d+1}(x-y)$. 
We will  then evaluate the general formula Eq.\,\eqref{eq:F1loopCas} in specific limits.
An arbitrary term of the series is represented in Fig.\,\ref{fig:diagrams}\textit{i}. 

\subsection{Vanishing of Tadpoles}
\label{se:tadpoles}

In Eq.\eqref{eq:F1loopCas}, each insertion of $\Sigma$ contains both $J_1$ and $J_2$. Let us focus on subterms involving only $J_2$. Such terms amount to the generalization of ``tadpole'' diagrams for an extended source, here $J_2$. 
Using  integrations by parts and the fact that the propagators $\Delta_0$ in empty spacetime are Lorentz invariant, \textit{i.e}  are functions of $u-v$ only, one can check that any such tadpole term  is  equal to minus itself and thus vanishes. This makes sense  since such terms do not involve the $J_1$ source at all and should not contribute to the force between the two sources. 
A tadpole diagram is represented in Fig.\,\ref{fig:diagrams}\textit{ii}. 
These diagrams contain  divergent contributions to the quantum work. Therefore the vanishing of the tadpole diagrams ensures that the perturbative finiteness  Eq.\,\eqref{eq:Finiteness1loop} is satisfied.

\subsection{The Scalar Casimir-Polder Limit}

Let us assume that the values of $\frac{n_{1,2}}{\Lambda}$  are small enough such that the leading contributions come from  the first terms in the series.
The first term of the series has $q=0$. This term amounts to a tadpole diagram, hence it vanishes as shown in Sec.\,\ref{se:tadpoles}.
We thus turn to  the $q=1$ term. This term is
\be
\W^{\rm 1-loop}_{q=1}=\frac{i}{2\Lambda^2} \int d^{d} {\bm u}\, d^{d+1} v\, \Delta_0(u,v)  J(v)   \Delta_0(v,u) \Lv \cdot \bpartial J_2(u)\,. 
\ee
We then decompose $J(u)=J_1(u)+J_2(u)$. The $\int J_2 \Delta_0^2  \bpartial J_2 $ piece is again a tadpole and thus vanishes as shown in Sec.\,\ref{se:tadpoles} --- using integration by parts, on can check that  it is  equal to minus itself. The remaining term is 
\be
\W^{\rm 1-loop}_{q=1} = \frac{i}{2\Lambda^2} \int d^{d} {\bm u} d^{d+1} v \, \Delta_0(u,v)  J_1(v)   \Delta_0(v,u) \Lv \cdot  \bpartial J_2(u) \,.
\label{eq:F1loopCP1}
\ee
Upon integrating by parts (or evaluating $\bpartial J_2$ and using the divergence theorem), we recognize the
variation of a bubble diagram that corresponds precisely to the definition of the Casimir-Polder potential $V_{\rm CP}(R)$ between two point sources. Namely, 
\be
\W^{\rm 1-loop}_{q=1} = -n_1n_2 \int d^{d} {\bm u} d^{d} {\bm v} \, \Lv\cdot \bpartial V_{\rm CP}(u-v) \,
\label{eq:F1loopCP2}
\ee
where we have defined the potential 
\be
 V_{\rm CP}(r) = -\frac{i}{2}\int dt \left(\Delta_0(0;r,t)\right)^2 = -\frac{1}{32\pi^3 \Lambda^2}\frac{m}{r^2}K_1(2 mr) \,.
\ee
We can see the explicit dependence on the fundamental mass $m$ in this result. 
Details of the explicit evaluation in the last step can be found in \textit{e.g.} Ref.\,\cite{Costantino:2019ixl}.

We can notice that Eq.\,\eqref{eq:F1loopCP2} amounts to the integral over the $J_{1,2}$ supports of the quantum work between two point sources generated by the directional derivative of the potential $W_{\rm CP}$, i.e. $\W^{\rm 1-loop}_{n=1} = n_1n_2 \int d^{d} {\bm u} d^{d} {\bm v} \,  W_{\rm CP}$ with $ W_{\rm CP} =  -\Lv\cdot \bpartial V_{\rm CP}$. 

A diagram in the Casimir-Polder limit is shown in  
Fig.\,\ref{fig:diagrams}\textit{iv}. In this limit the quantum loop penetrates the whole bodies.

\subsection{The  Scalar Casimir Limit}
\label{se:casimir_limit}

A different  limit is obtained when  the effective mass inside the sources, $m^2+\frac{n_{1,2}(\x)}{\Lambda}$,
is large enough for the dressed propagator to be repelled from the sources.
This occurs whenever ${\cal D}_x \Delta(x,x')\ll J(\x)\Delta_J(x,x')$ for any $x, x'$. 
In this Dirichlet limit, 
the EOM Eq.\,\eqref{eq:EOMgen} gives then $J(\x)\Delta_J(x,x')\approx  0$, which enforces $\Delta_J(x,x')\approx 0$ for any $x$ in the source and any $x'$ in the whole space. The propagator vanishes on the boundary,  $\Delta_J(x\in \partial J,x')\approx 0$, by continuity. Therefore the propagator has Dirichlet boundary conditions in this regime. 
 We refer to this limit as the scalar Casimir limit since it reproduces a Dirichlet problem for which the quantum force is usually referred to as "Casimir" even when the underlying theory is not electrodynamics, e.g. here a massive scalar field theory. 
 
A sample diagram from the Casimir limit is shown in
Fig.\,\ref{fig:diagrams}\textit{iii}. 
{In this limit  the quantum loop does not penetrate inside the  bodies. }

Summarizing, we have shown that our formula for the one-loop quantum work Eq.\eqref{eq:F1loopCas} interpolates between the scalar Casimir-Polder force and the scalar  Casimir force.  The two limits are realised in different physical regimes, which essentially depend on the competition between the magnitudes of the effective mass and of  the d'Alembertian in the EOM. Qualitatively, we can say that for fixed fundamental parameters and densities, the Casimir-Polder limit emerges in the
\textit{short separation} regime while the Casimir limit emerges in the  \textit{large separation} regime.

\section{The  Dirichlet Sphere }
\label{se:Sphere}

In this section we consider the  Casimir pressure on a spherical shell in the presence of  a scalar field  with Dirichlet boundary condition on the shell, \textit{i.e.} a ``Dirichlet sphere''. This is a standard problem, that we wish here to revisit.  An early calculation can be found in  \cite{Boyer:1968uf}, and an 
expression describing  the Casimir pressure on a  $d-1$-sphere   has been derived in \cite{Milton_Sphere}, which will be our  main reference.

\subsection{Is the quantum pressure on the  sphere  finite in QFT?}

\label{se:why}

The QFT prediction obtained in \cite{Milton_Sphere} features a ``spurious'' divergence. 
Here we  discuss why such a divergence is not expected in light of the  finiteness properties derived in Sec.\,\ref{se:quantumwork}. 
We have  emphasised in Sec.\,\ref{se:quantumwork}  the role of matter conservation to obtain a finite quantum work.  In the case of the sphere, the deformation flow describing the radial deformation of the sphere is not divergence-free, $ \bpartial \cdot L_{\rm Sphere} \neq  0$ for arbitrary spacetime dimension $d$ except $d=1$. If the sphere density were to be  assumed to be constant, then the matter of the sphere would \textit{not} conserved under the deformation and it would then  follow that neither of the finiteness properties \eqref{eq:Finiteness} and  \eqref{eq:Finiteness2} would apply. Hence a divergent piece would show up in the expression of the quantum pressure as an artefact. This is exactly what we find below. We also show how this spurious  divergence exactly cancels when matter conservation is used.

\subsection{Review and Discussion}

We first review the result from \cite{Milton_Sphere}.
 The radius of the sphere is denoted by $a$.  The pressure on the sphere obtained in this reference can be put in the form
 \be
 \frac{F_{S_{d-1},[\color{blue}{26}]}} {A_{d-1}} =  \frac{F^{\rm fin}_{S_{d-1},[\color{blue}{26}]}}{A_{d-1}} +  \frac{F^{\rm div}_{S_{d-1},[\color{blue}{26}]}}{A_{d-1}}
 \label{eq:FSphere}
 \ee
 where
\begin{eqnarray}
&&\frac{F^{\rm fin}_{S_{d-1},[\color{blue}{26}]}}{A_{d-1}}  = i \frac{1}{a^{d}} \sum^\infty_{h=0} c_h \int^\infty_{-\infty} d\omega  \omega \frac{d}{d\omega} \log\left( \omega a J_{h-1+\frac{d}{2}}(|\omega| a)H^{(1)}_{h-1+\frac{d}{2}}(|\omega| a)\right)  
 \label{eq:FSphereFin}\nonumber \\
&&
\frac{F^{\rm div}_{S_{d-1},[\color{blue}{26}]}}{A_{d-1}} =  i\frac{1-d}{a^{d}} \sum^\infty_{h=0} c_h \int^\infty_{-\infty} d\omega.
 \label{eq:FSphereInf}
\nonumber \\
\end{eqnarray}
and the coefficients are given by
\be 
c_h = \frac{(h-1+\frac{d}{2})\Gamma(h+d-2))}{2^{d}\pi^{\frac{d+1}{2} } h!\Gamma(\frac{d-1}{2})}.
 \label{eq:cn}
\ee
$A_{d-1}$ is the area of the $d-1$-sphere with radius $a$. All the terms are real upon rotation to Euclidean time, here for convenience we keep the Lorentzian integrals. 
The result Eq.\,\eqref{eq:FSphere} is obtained based on the difference between the radial component of the  stress tensor on each side of the sphere.  In our language, this is equivalently obtained by considering the sphere as a rigid source with infinite density and deforming it along the radial flow $\Lv = {\bm e}_r$.

The $F_S^{\rm fin}$ term is finite for any $d$ different from positive even integers. The divergence showing up when $d=2k$, $k=1,2\ldots$  is a familiar  feature in QFT that can be treated in the framework of renormalization. This physical divergence is not our focus here. In contrast, the $F_S^{\rm div}$ is infinite for any $d\neq 1$. This behavior is not the one of a meromorphic function in $d$, and should thus draw our attention.

In   \cite{Milton_Sphere} it was nicely observed that for $d<1$ the $\sum^\infty_{h=0} c_h $ series vanishes identically. A proposal was then made to remove the $F^{\rm div}$ term using an analytical continuation of $d$ to the $d<1$ region. In summary the argument amounts to stating that since in this region one has $\sum^\infty_{h=0} c_h =0$ identically,  any term constant in $h$ arising from the integral is irrelevant since it is multiplied by zero and can thus be subtracted. 
Notice that the proposed argument is different from  usual dimensional regularization which simply turns the quantity of interest into a meromorphic function of $d$. 

An issue {remains}, however, { as }  $\sum^\infty_{h=0} c_h =0$ in the $d<1$ region {only guarantees that} the divergent term 
$F^{\rm div}$ takes the indefinite form ``$0\times \infty$''. As a result, even if one requires the $\sum^\infty_{h=0} c_h$ sum to vanish for any $d>1$ by analytical continuation of zero,  the proposed method  remains inconclusive in the sense that $0\times \infty$ is undefined.  {For this reason, we will see that the use of our formalism (namely the quantum work on a Dirichlet shell Eq.\,\eqref{eq:Wshell2} together with matter conservation)  allows one to bypass these ambiguities and make the result finite.}

\subsection{The Quantum Work on the Dirichlet Sphere}

We now proceed with our  calculation of the quantum work on the Dirichlet sphere with radial deformation. We first define the source and the deformation.
We consider a source with the  geometry of a spherical shell of width $\eta$ and with a finite number density $n$ in the $\eta\to 0$ limit, 
\be J_\eta(\x) = \frac{n_\lambda}{\eta} {\bm 1}_{a-\frac{\eta}{2} <r<a+\frac{\eta}{2}}(\x)\,\ee 
The propagator in the presence of this source is denoted by $\Delta_S(x,x')$.
The boundary of the shell for small $\eta$ is identified with  $\partial S_{\eta\to 0} = S_{\rm in}\cup S_{\rm out}$ where $S_{\rm in,out}$ are the $d-1$-spheres with respective radii $r=a_{-},a_{+}$. 
The deformation of the source is parametrized by $\lambda$ and changes the sphere radius such that $a_{\lambda+d\lambda}= a_{\lambda}+  L d\lambda $. Equivalently, in terms of the support function, the deformation is $l_{\lambda+d\lambda}(r)=l_{\lambda}(r-L d\lambda)$.  
The deformation flow vector is thus $\Lv=L {\bm e}_r$. 
Using the conservation equation Eq.\,\eqref{eq:cons_gen} we can easily derive the variation of density corresponding to such a deformation. We find
\be
\partial_\lambda n= -\frac{d-1}{a} n L \,. \label{eq:deformation_sphere}
\ee
We can now compute the quantum work. Since we are interested in a sphere we can readily use the general formula for the  quantum work on a thin shell Eq.\,\eqref{eq:Wshell2}.  In the Dirichlet limit the quantum work  reads
\begin{align}
& W_{S_{d-1}}= \nn \\ \nn
&
-\frac{1}{2}A_{d-1} L \left[\partial_{r'}  \partial_{r''} \DeltaD_S(r',t;r'',t)|_{r'=r''\in S_{\rm out}}  -
\partial_{r'}  \partial_{r''} \DeltaD_S(r',t;r'',t)|_{r'=r''\in S_{\rm in}} \right] \nn
\\ &
-\frac{1}{2\Lambda} \int_{S} d^d\x  \Delta_S(x,x) \partial_\lambda n_\lambda \Big|_{n\to\infty} \,. 
\label{eq:Sphere1}
\end{align}
The first term in Eq.\,\eqref{eq:Sphere1} matches  precisely the quantity computed in \cite{Milton_Sphere}. Namely we find
\begin{align}
 W_{S_{d-1}}= 
L\left(F^{\rm fin}_{S_{d-1},[\color{blue}{26}]}+  F^{\rm div}_{S_{d-1},[\color{blue}{26}]}\right) 
-\frac{1}{2\Lambda} \int_{S} d\sigma(\x)  \Delta_S(x,x) \partial_\lambda n_\lambda \Big|_{n\to\infty} \,
\label{eq:Sphere2}
\end{align}
where the components of the force are defined in Eqs.\,\eqref{eq:FSphereFin}. 

The remaining task is to evaluate the last term in Eq.\,\eqref{eq:Sphere2}, which encodes  the variation of density. 
To this end we first have to evaluate the propagator in the presence of the sphere with finite density. This is done by recomputing the propagator in \cite{Milton_Sphere}, replacing the two Dirichlet boundary  conditions on $S$ by two boundary conditions obtained from integrating the EOM on the  shell enclosing $r=a$ and using the divergence theorem. 
Introducing the Fourier transform in time $\Delta(x,x)=\int \frac{d\omega}{2\pi} \Delta_\omega(\x,\x)$, we find for the propagator at coinciding points
\be
\Delta_\omega(a,a) \overset{{\rm large} \,n } {\to} i \frac{\Lambda}{n a^{d-1}}\sum^{\infty}_{h=0}  \frac{(h-1+\frac{d}{2})\Gamma(h+d-2))}{2^{d-2}\pi^{\frac{d-1}{2} } h!\Gamma(\frac{d-1}{2})}
= i \frac{\Lambda}{n a^{d-1}} \sum^{\infty}_{h=0} 4\pi c_h 
\ee
with $c_h$ defined in Eq.\,\eqref{eq:cn}. 
Using the variation of density dictated by matter conservation Eq.\,\eqref{eq:deformation_sphere} we then obtain
\be
\frac{1}{2\Lambda} \int_{S} d\sigma(\x) \Delta_{S}(x,x) \partial_\lambda n_\lambda \Big|_{n\to\infty} = i\frac{1-d}{a^d}  L \sum^\infty_{h=0} c_h \int^\infty_{-\infty} d\omega  =L F^{\rm div}_{S_{d-1},[\color{blue}{26}]}\,.
\ee
We see that this contribution from the variation of  density exactly cancels the divergent piece in Eq.\,\eqref{eq:Sphere2}. It follows that our final result for the quantum work on the Dirichlet sphere amounts to the finite part of the result from \cite{Milton_Sphere}, namely 
\be
 W_{S_{d-1}}= L F^{\rm fin}_{S_{d-1},[\color{blue}{26}]}\,.
\ee

The fact that the term from the variation of density cancels the $F^{\rm div}_{S_{d-1},[\color{blue}{26}]}$ divergence upon requirement of matter conservation is  non trivial.  This cancellation provides a check of our expression for the  quantum work on a thin shell Eq.\,\eqref{eq:Wshell2} and illustrates how  the finiteness of the quantum work can manifest itself concretely.

\section{Planar Geometry}
\label{se:planar}

In this section we evaluate the quantum force in simple planar geometries. While the geometric setup in itself is very well-known, the exact results  for the specific quantum force considered here in the case of a massive scalar have  not been presented elsewhere. This detailed calculation  serves to illustrate how the quantum work remains finite as a result of matter conservation. It also exhibits the transition between the scalar Casimir and Casimir-Polder regimes. The plate-point result obtained here is also key for the search for new particles via atom interferometry  that is presented in   section \ref{se:interferometry}.

\subsection{Force Between two Plates}
\label{se:plane-plane}

We focus  on the classic Casimir setup with two  plates facing each other and separated by a distance $\ell$ along the $z$ axis. The deformation we consider amounts to a variation of $\ell$. 
We compute the force  induced by a massive scalar field with bilinear coupling to the constituents of the plates.

The quantum field $\Phi$ is described by the Lagrangian 
\be
{\cal L}=\frac{1}{2}(\partial_\mu \Phi)^2 -\frac{m^2}{2}\Phi^2 - \frac{1}{2\Lambda^2}\Phi^2J( {\bf x })\,. \label{eq:Lplaneplane}
\ee
In this application, it is convenient  to define $J$ to be the mass density distribution of the sources.  
In this setting
there are   five  regions along $z$, with the  plates supported on region $1$ and $3$. The source $J$ is defined as
\be
J(z)=    \rho_1  \Theta(-\bar z_\infty<z<0)
+\rho_3\Theta(\ell<z<z_\infty)  \label{eq:sourceplaneplane}
\ee
The width of the plates is taken to be much larger than the separation, $z_\infty,\bar z_\infty\gg \ell$. The fact that the plates actually end instead of continuing to infinity \textit{i.e.} $|z_\infty|,|\bar z_\infty|<\infty$ is crucial in order to ensure matter conservation, and therefore that the quantum work is finite as dictated by Eq.\,\eqref{eq:Finiteness2}.  The effective mass can be written as
\begin{align}  & m^2(z)= m_\infty^2 \Theta(z<-\bar z_\infty)+   m_1^2  \Theta(-\bar z_\infty<z<0)+m_2^2 \Theta(0<z<\ell)
\nonumber \\
&+m_3^2 \Theta(\ell<z<z_\infty) + m_\infty^2 \Theta(z>z_\infty)  \,, \end{align}
where 
\be 
m_{\infty,2}^2= m^2
\ee
and 
\be 
m_{1,3}^2= \frac{\rho_{1,3}}{\Lambda^2}+m^2\,. \label{eq:m13planeplane}
\ee
In Eq.\,\eqref{eq:Lplaneplane} the boundary operator amounts to 
\be 
{\cal B}_m(\Phi)= \frac{\Phi^2}{2\Lambda^2}\,.
\ee
 Derivative operators such as ${\cal B}(\Phi)= \frac{(\partial \Phi)^2}{2\Lambda^4}$ could be treated along the same lines.

The deformation of the source that we consider amounts to shifting  the right plate (\textit{i.e.} region 3,  the second term in Eq.\,\eqref{eq:sourceplaneplane}.  
This corresponds to an  infinitesimal shift of $\ell$ and $z_\infty$, $\ell_{\lambda+d\lambda}=\ell_\lambda +Ld\lambda $, $z_{\infty,\lambda+d\lambda}=z_{\infty,\lambda} +Ld\lambda $.
Notice that the left plate is left untouched, hence $\bar z_\infty$ does not vary.  
Equivalently, in terms of the support function of the right plate, this is described by $l_{\lambda+d\lambda}(z)=l_{\lambda}(z- L d\lambda)$. The geometry is summarized as
\be
\includegraphics[width=0.8\linewidth,trim={0cm 9cm 0cm 5cm},clip]{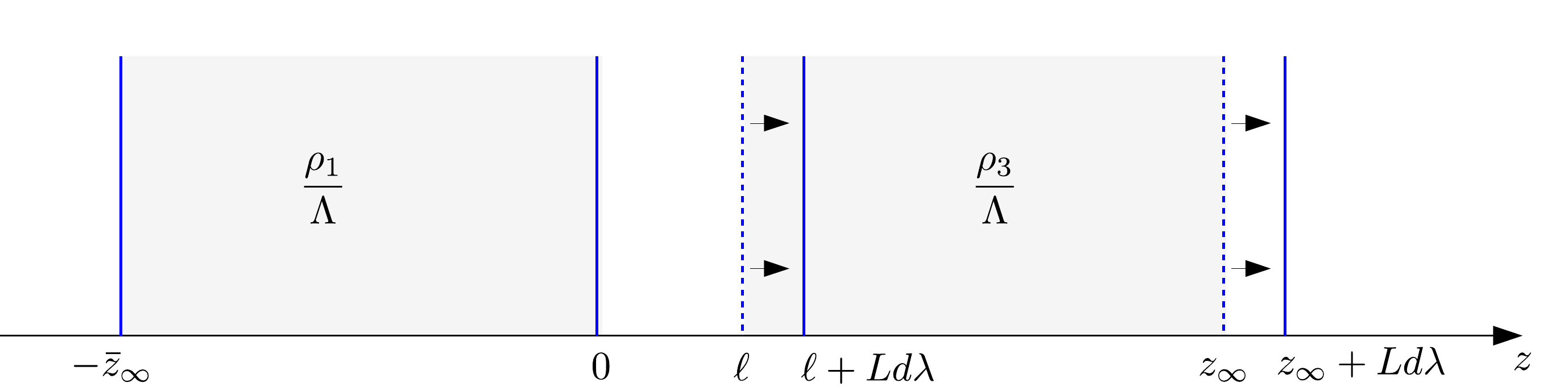} \nonumber
\ee
Since the source moves rigidly, the formula for the quantum work Eq.\,\eqref{eq:F1loopCas} applies and a quantum force can be defined out from the quantum work, $W^{\rm 1-loop}= L F_{\rm quant}$.  In the following we determine $F_{\rm quant}$.

\subsubsection*{Propagator}

The Feynman propagator in  position-momentum space, defined by 
\be 
\Delta(x,x')=\int\frac{d^3p}{(2\pi)^3}e^{ip^\alpha(x-x')_\alpha} \Delta_p(z,z')
\ee
with  $(p^\alpha, z)$, $\alpha=(0,1,2)$, 
has been calculated in the presence of a piece-wise constant mass in  \cite{Brax:2018grq}.
Defining the $z$-dependent momentum with the Feynman $\epsilon$-prescription \be 
\omega(z)=\sqrt{(p_0)^2-(p_1)^2-(p_2)^2+i \epsilon-m^2(z)},
\ee
the homogeneous equation of motion becomes
\be(\partial_z^2 + \omega^2(z)) \Phi(z)=0\ee
whose solutions in a given region $i$ are simply  $e^{\pm i\omega_i z}$. The solution everywhere   can be found by  continuity of the solution and its derivative at each of the  interfaces. The propagator is obtained by solving the equations of motion in the  five regions and matching them at the boundary. Details can be found in the appendix of \cite{Brax:2018grq}. 
For the calculation of the quantum force, we will only need to evaluate the propagator at coinciding points on the two boundaries of the right-hand plate, $z=\ell$ and $z=z_\infty$.

\subsubsection*{Quantum Force}

The deformation of the source term is found to be 
\be
\partial_\lambda J =  - (\rho_3-\rho_2) L \left(\delta(z-\ell) -\delta(z-z_\infty)\right) \,.
\ee
Putting  this variation back into the definition of the quantum work, we can factor out the $L$ term and obtain the  quantum force as defined in Eq.\,\eqref{eq:F1loopCas}. 
As a result the quantum force is given by
\begin{align}
{ F}_{\rm quant}  & =  \frac{1}{2}(m_3^2-m_2^2) \int d^2{\bm x}_\parallel ( \Delta(x^\alpha,\ell;x^\alpha,\ell)-\Delta(x^\alpha,z_\infty;x^\alpha,z_\infty))
\nn \\ & =  \frac{1}{2}(m_3^2-m_2^2) \int d^2{\bm x}_\parallel \int \frac{d^3p}{(2\pi)^3}  ( \Delta_p(\ell,\ell)-\Delta_p(z_\infty,z_\infty))
\label{eq:Fquant1}
\end{align}
with ${\bm x}_\parallel = (x_1,x_2)$. 
The cancellation  between the divergent parts of the two propagators at coinciding points is evident in Eq.\eqref{eq:Fquant1}, using the fact that  the divergence is location-independent. In the second line we have introduced the propagator in position-momentum space. 
Here we have explicitly 
\be
\Delta_p(\ell,\ell)-\Delta_p(z_\infty,z_\infty)=
\frac{  (\omega_1+\omega_2)+  e^{2i \ell \omega_2} (\omega_2-\omega_1)
}{ (\omega_1+\omega_2)
(\omega_2+\omega_3)-
e^{2i \ell \omega_2} (\omega_2-\omega_1)(\omega_2-\omega_3)
}-\frac{1}{\omega_2+\omega_3}\,.
\ee
with $\omega_i = \sqrt{(p_\alpha)^2+i\epsilon-m^2_i}$. 

The surface integral $\int d^2 {\bm x}_\parallel=S$  is factored out hence defining a pressure.  The final expression for the quantum pressure between the two plates (\textit{i.e.} regions $1$ and $3$) is then
\be
\frac{F_{\rm quant }}{S} =
 \int_0^\infty \frac{d \q \q^2}{2\pi^2} 
 \frac{ \gamma_2 (\gamma_2-\gamma_1)(\gamma_2-\gamma_3)
}{
 (\gamma_2-\gamma_1)(\gamma_2-\gamma_3)
 -
e^{2 \ell \gamma_2}
 (\gamma_1+\gamma_2)
(\gamma_2+\gamma_3)
}\, \,
\label{eq:Density_cham}
\ee
after  Wick's rotation with $\omega_i=i\gamma_i=i \sqrt{\q^2+m^2_i}$. This is the general expression of the plate-plate quantum pressure in the presence of a scalar coupled quadratically to matter.

In the limit of large density-induced effective mass  $m_{1,3}\rightarrow \infty$, the general expression Eq.\,\eqref{eq:Density_cham} becomes
\be
\frac{F_{\rm quant }}{S} =  \int_0^\infty \frac{d \q \q^2}{2\pi^2}
 \frac{\gamma_2
}{
1- e^{2 \ell \gamma_2} 
}\,. \label{eq:Density_cham_Cas}
\ee
In this limit  the effective mass in the plates become so large that the field obeys Dirichlet boundary conditions. 
Accordingly, Eq.\,\eqref{eq:Density_cham_Cas} matches exactly the Casimir pressure from a massive scalar {{with Dirichlet boundary conditions}}. 
For a massless scalar the integral can be explicitly performed and we retrieve
\be 
\frac{F_{\rm quant }}{S}= -\frac{\pi^2}{480\ell^4} \,. 
\ee
 This is the classic Casimir pressure for a massless scalar field. 
 
In the limit of small density-induced effective mass defined as $(m^2_{1,3}-m_2^2)/ m^2_2\ll 1$, \textit{i.e.}
when the contribution of the density to the effective mass is small with respect to the fundamental mass, the pressure becomes
\be
\frac{F_{\rm quant }}{S} = -  (m^2_{1}-m_2^2)(m^2_{3}-m_2^2)
\int_0^\infty \frac{d \q \q^2}{2\pi^2}
 \frac{e^{-2 \ell \gamma_2}
}{16
 (\gamma_2)^{3} }\,. \label{eq:Density_cham_CP}
\ee
We checked that this  corresponds exactly to the Casimir-Polder force integrated over regions 1 and 3. We will see how this calculation can be cross-checked in the next example. 

In summary, we have verified that both the scalar Casimir and scalar Casimir-Polder pressures are recovered as limits of the more general expression of the plate-plate quantum pressure  Eq.\,\eqref{eq:Density_cham}. Qualitative considerations are given in the next example.

\subsubsection*{On Finiteness}

In Eq.\,\eqref{eq:Fquant1} we have observed the cancellation between the divergent piece of  $\Delta_p(\ell,\ell)$ and $\Delta_p(z_\infty,z_\infty)$ in the integrand. This cancellation makes  the expression for the force finite. 
To illustrate how the presence of the $\Delta_p(z_\infty,z_\infty)$ term is tied to matter conservation, we consider the following counterexample. 

Let us imagine that we ignored the displacement of the outer edge ($z=z_\infty$) of the plate. This would imply that the matter of the plate is not conserved since the plate's width would  change while the density remains constant. 
In such a setup, the result would be the same as Eq.\,\eqref{eq:Fquant1} but without the $\Delta_p(z_\infty,z_\infty)$ contribution. Therefore the expression would be  infinite. This simple counterexample illustrates that, when dropping the requirement of matter conservation, the prediction of the force becomes infinite \textit{i.e.} property \,\eqref{eq:Finiteness2} does not hold.

\subsection{Force Between a Plate and a Point Source }
\label{se:point-plane}

As a third application of our formalism, we  focus on the interaction between a point particle and a plate. 
 As before we assume the Lagrangian 
\be
{\cal L}=\frac{1}{2}(\partial_\mu \Phi)^2 -\frac{m^2}{2}\Phi^2 - \frac{1}{2\Lambda^2}\Phi^2\,\, J( {\bf x })\,.
\ee
The plate is supported on $z<0$  and has mass density $\rho_1$. 
The source is taken to be
\be
J( {\bf x })= \rho_1 \Theta(-z_\infty<z<0) +  m_N { \delta^2 }({\bf x_\parallel})\delta(z-\ell)\,.
\ee
The mass of the point particle is $m_N$. 
We define the effective mass of $\Phi$ in the plate as 
\be
 m^2_1 = \frac{\rho_1}{\Lambda^2} +m^2 \,
\ee
which  depends on the coupling to matter, $\frac{1}{\Lambda^2}$.
The effective mass of $\Phi$ is then piecewise constant,
\be
m^2(z)=m_\infty^2 \Theta (z<-z_\infty)+ m_1^2\Theta(-z_\infty<z<0)+m_2^2\Theta(z>0)
\ee
with $m_{\infty,2}=m$.

The deformation  we consider is an infinitesimal  shift of the point particle position $\ell$, $\ell_{\lambda+d\lambda}=\ell_{\lambda}+L d\lambda$. 
The geometry is summarized as
\be
\centering
\hspace{3cm }\includegraphics[width=0.8\linewidth,trim={0cm 9cm 0cm 5cm},clip]{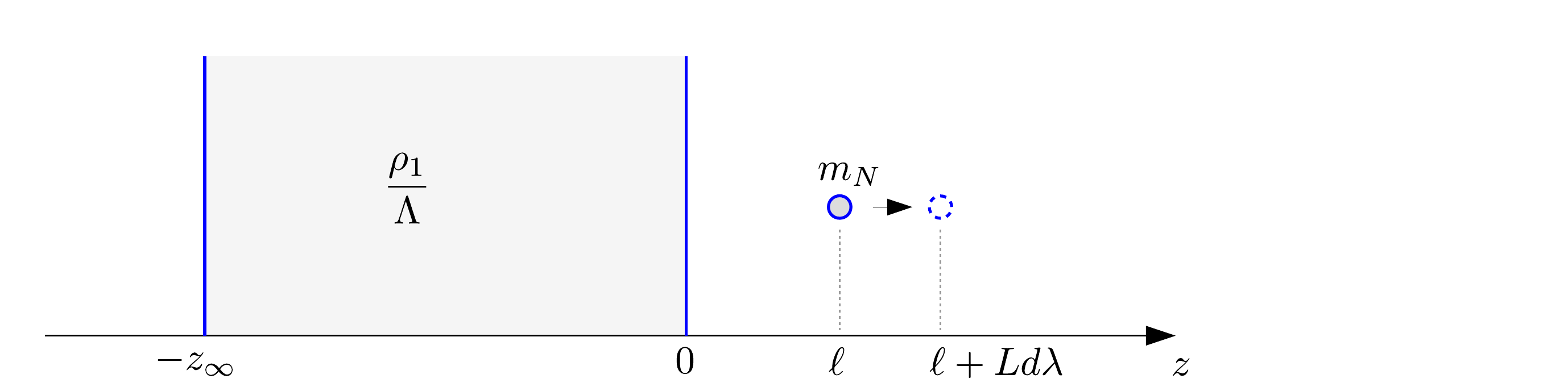} \nonumber
\ee

\subsubsection*{Propagator}

The Feynman propagator in  position-momentum space $(p^\alpha, z)$, $\alpha=(0,1,2)$ in the presence of a piecewise constant mass  has been calculated in  \cite{Brax:2018grq}.
The effect of the point source on the propagation is negligible.\,\footnote{This can be checked by evaluating the dressed propagator in energy-position space $(p_0,\x)$.
In the resummed propagator, the effect of the insertion is small within the EFT validity range, leaving the term with one point source insertion as the main non-vanishing contribution to the quantum work. } 
For the present calculation, if the frame is chosen such that the deformation changes  the position of the point source and not of the plate, one can safely ignore the region at $-z_\infty$ and thus consider the propagator over two regions, with $m^2(z)=m_1^2\Theta(z<z_{12})+m_2^2\Theta(z>z_{12})$. 
The propagator is found to be
\be
\Delta_p(z, z')= \label{eq:propa2regions}
\begin{cases}
 \frac{e^{i\omega_2 (z_>-z_<)}}{2\omega_2}E_2(z_<)   \quad\quad z_{12}<z_<
 \\
 \frac{e^{i(\omega_2 (z_>-z_{12})-\omega_1 (z_<-z_{12})}}{\omega_1+\omega_2}   \quad z_<<z_{12}< z_>
  \\
  \frac{e^{i\omega_1 (z_>-z_<)}}{2\omega_1}E_1(z_>)  \quad\quad  z_><z_{12}
\end{cases}\,
\ee
where
\begin{eqnarray}
&& E_1(z)=1+ e^{i2(z_{12}-z)\omega_1}\frac{\omega_1-\omega_2}{\omega_1+\omega_2}\nonumber \\ && E_2(z)=1+ e^{i2(z-z_{12})\omega_2}\frac{\omega_2-\omega_1}{\omega_1+\omega_2}\,. 
\nonumber \\
\end{eqnarray}
We have defined $z_<= \min (z,z')$ and $z_>=\max (z,z')$. We have introduced $\omega_i=\sqrt{(p_\alpha)^2-m^2_i +i\epsilon}$. The $\epsilon$ prescription guarantees that the propagators decay  at infinity. 
The $E_1$, $E_2$ functions essentially describe how the presence of the boundary affects the propagator with both endpoints in the same region. When the boundary $z_{12}$ is rejected to infinity, one recovers the usual expression for  a fully homogeneous space.

\subsubsection*{Quantum Force}

The deformation of the source is 
\be
\partial_\lambda J =  -m_N L { \delta^2 }({\bf x_\parallel})\partial_z\delta(z-\ell) \,.
\ee
Using this expression into the quantum work, one obtains after one integration by parts the quantum force
\begin{align}
{ F}_{\rm quant}   & =  -
 \frac{1}{2}\frac{ m_N}{\Lambda^2}   \partial_z\Delta_J(x^\alpha,z;x^\alpha,z)|_{z\to\ell}
 \\ \nn &  =  -
 \frac{1}{2}\frac{ m_N}{\Lambda^2}   \int\frac{d^3p}{(2\pi)^3}    \partial_z\Delta_J({\bm k};z,z)|_{z\to\ell}
 \end{align}
 In the last line we have introduced the position-momentum space propagator.

Using Eq.\,\eqref{eq:propa2regions}   with $z_{12}=0$  since the plate is placed at the origin, we have
   \begin{align}
 F_{\rm quant}  =  & -\frac{ m_N}{\Lambda^2} 
 \frac{1}{2} \int\frac{d^3{ p}}{(2\pi)^3}   \frac{ 1}{2\omega_2} \partial_z E_2(z)|_{z=\ell}
\nn \\  = & 
-(m_1^2-m_2^2)\frac{ m_N}{\Lambda^2} \frac{1}{4\pi^2}  \int d\q \,\q^2    \frac{e^{ - 2 \ell\gamma_2}}{(\gamma_1+\gamma_2)^2}\,
\label{ppo}
   \end{align}
where  we have  performed a Wick rotation and introduced $\gamma_i=\sqrt{\q^2+m^2_i}$. 
This is the general expression of the plate-point quantum force in the presence of a scalar coupled quadratically to matter. 
Diagrammatically,  the particle will interact with the plate via loops starting at the point particle, going into the plate and coming back to the point particle.

In the limit of  large density-induced effective mass $m_1\to\infty$, the force takes the form
\be
 F_{\rm quant}    = - \frac{ m_N}{\Lambda^2} \frac{1}{4\pi^2}  \int d\q \,\q^2   e^{ - 2\ell \gamma_2} \,.
\ee
This is the limit for which the density  is so large that the field is repelled by the plate and the field obeys Dirichlet boundary conditions at the boundary of the plate. We refer to this case as the Casimir limit. 
In the massless case we obtain
\be
 F_{\rm quant}  = -\frac{m_N}{16\pi^2 \Lambda^2}\frac{1}{\ell^3} \,.
  \label{eq:FCmassless}
\ee

In the limit of small density-induced effective mass  we expand in $(m^2_1-m_2^2)\ll m_i^2$ and obtain
\be
 F_{\rm quant}    = - (m_1^2-m_2^2)\frac{ m_N}{\Lambda^2} \frac{1}{4\pi^2}  \int d\q \,\q^2   e^{ - 2 \ell\gamma_2} \frac{ 1}{4 \gamma_2^2 } \label{eq:CP}
\ee
In the massless case we obtain
\be
 F_{\rm quant}    =  -\frac{m_1^2\,m_N}{16\pi^2 \Lambda^2}\frac{1}{\ell}\,.
 \label{eq:FCPmassless}
\ee
This is the Casimir-Polder limit. As a cross check, in the Appendix we show that this limit is exactly recovered by integrating  the point-point Casimir-Polder potential over the extended source.

In summary, we have verified that both the scalar Casimir and scalar Casimir-Polder pressures are recovered as limits of the more general expression of the plate-point quantum pressure  Eq.\,\eqref{eq:Density_cham}.
The two limits of the  massless formula Eqs.\,\eqref{eq:FCmassless} and \eqref{eq:FCPmassless}  make transparent that there is a transition between the two regimes as a function of the separation $\ell$. Namely, the Casimir regime occurs for $\ell\gg m^{-1}_1 $ while the Casimir-Polder regime occurs for $\ell\ll m^{-1}_1 $, with $m^2_1=\frac{\rho_1}{\Lambda^2}$. One way to think about this phenomenon is that, while at large distance the plate behaves as a mirror, leading to a Casimir force, at short distance the quantum fluctuations start penetrating the mirror. As a result the behaviour of the Casimir force gets softened into the the Casimir-Polder one at short distance. 
This behaviour is confirmed numerically.

\section{Bounding Quantum Forces with Atom Interferometry}

\label{se:interferometry}

\subsection{The setting}

In this section we calculate a somewhat more evolved observable in the context of  atom (or neutron) interfero\-metry experiments. Namely we compute the matter-wave phase shift  measurable in  atom interferometers, which is generated in the presence of a quantum force between the atom and a neighbouring plate. The calculation uses the results from section \ref{se:point-plane}.

Atom interferometry has been used to test gravity (see \textit{e.g.} \cite{KasChu, KasChu2,1999Natur.400..849P,2006PhyS...74C..15M}) and to search for classical fifth forces, often in the context of dark energy-motivated models \cite{Burrage:2014oza,Hamilton:2015zga,Jaffe:2016fsh,Sabulsky:2018jma}. 
Interferometry has never been used to search for quantum forces such as the one modelled by the scalar field used throughout this paper.  In this section we thus \textit{i) }compute the phase shift induced by a quantum force and \textit{ii)} demonstrate that interferometry is a competitive method to search for a dark field bilinearly coupled to matter. 

\begin{figure}
\centering
			\includegraphics[width=0.8\linewidth,trim={2cm 2cm 4cm 3cm},clip]{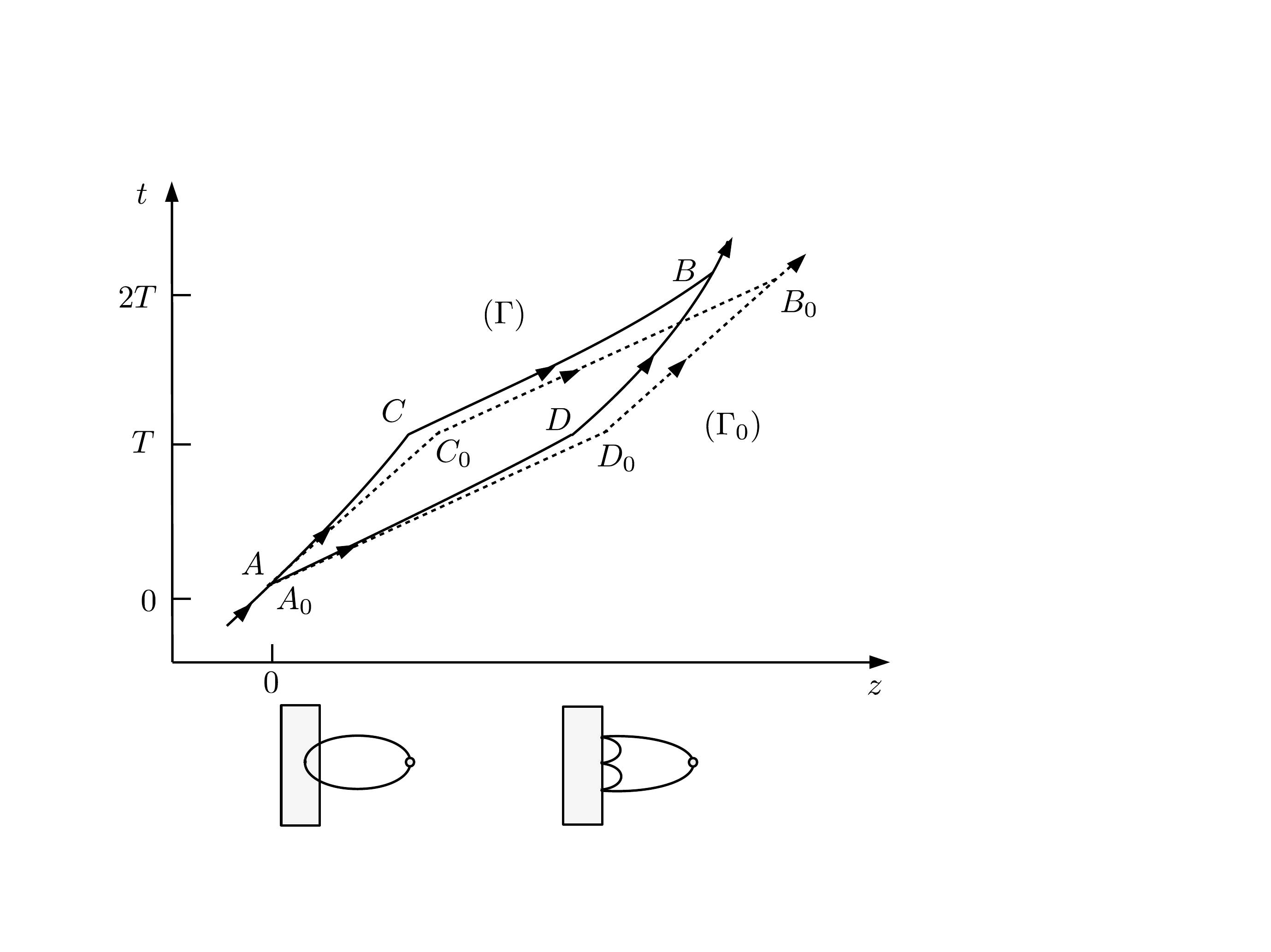}
\caption{ Spacetime paths followed by the atoms in the interferometer. The path in the presence of the force ($\Gamma$, plain lines) is deformed with respect to the path in the absence of the force ($\Gamma_0$, dotted lines). The pictures indicate the asymptotic behaviors (Casimir-Polder and Casimir) of the force between the atom and the plate. 
\label{fig:diags_interferometry}}
\end{figure}

Our focus is on a setup which is essentially an adaptation of the simplest ``Kasevich-Chu'' experiment (see \cite{KasChu, KasChu2,1999Natur.400..849P}, and \textit{e.g.}  \cite{2006PhyS...74C..15M} for a review).       We describe briefly the setup and  refer to the above references and to \textit{e.g.} \cite{Storey:1994oka} for further details.  Atom or neutron  interferometry uses the difference of phase of two coherent wavepackets following two different spacetime paths as shown in Fig.\,\ref{fig:diags_interferometry}.
The two paths amount to  two broken  worldlines, $ACB$ and $ADB$, with a change of direction at $C$ and $D$ respectively, and with same endpoint $B$.  The changes of velocity of the wavepackets are induced by laser pulses.

We  assume that the interferometry experiment is carried out along the $z$ axis over a plate located at $z=0$. We assume that the setup is oriented horizontally at the surface of the Earth since we are not interested in measuring the strength of the gravity field. 
The phase shift is caused by the  force between the atom and the plate.   Experimentally, the plate might be a large ball whose radius is assumed to be much larger than  the length of $\Gamma$. 
In this setup the potential $V(z)$ is the one for  the plane-point quantum force derived in section \ref{se:point-plane}.

Using the WKB approximation, the leading phase shift between the two paths induced by a potential $V(z)$  is given by (see \cite{Storey:1994oka})
\be
\delta \phi = - \int_{\rm \Gamma_0} dt\  V(z(t))\
\ee
where $\Gamma_0=A_0C_0B_0D_0$ denotes the unperturbed closed path  and $z(t)$ the corresponding classical trajectory of the wavepackets. 
Along each segment of $\Gamma_0$, the trajectories are straight, 
\be
z(t)= z_{i}+v_i(t-t_i)\,.
\ee
 The velocities along each segment of the path can in general differ.  In the present setup the time spacing between each pulse is  $T$ and the velocities on the segments of $\Gamma_0$ are $v_{A_0C_0}=v_{D_0B_0}=v$, $v_{C_0B_0}=v_{A_0D_0}=v'$, as shown in Fig.\,\ref{fig:diags_interferometry}.

\subsection{Computing the Phase Shift}

We consider the scalar model defined in Eq.\,\eqref{se:point-plane}, where the fundamental mass of the field is denoted by $m$ and the mass density of the plate is denoted $\rho$.  The effective mass of the $\Phi$ field  for $z<0$ and $z>0$ is given by $m_1^2=\frac{\rho}{\Lambda^2}+m^2$ and $m_2^2=m^2$. 
The plane-point force (\ref{ppo}) derives from a potential $F(\ell)= -\frac{\partial V(\ell)}{\partial \ell}$ given by
   \be
 V(\ell)    =   -\frac{ m_N\rho }{\Lambda^4} \frac{1}{4\pi^2}  \int d\q \,\q^2    \frac{e^{ -  2 \ell\,\gamma_2}}{2 \gamma_2(\gamma_1+\gamma_2)^2}\, \label{eq:V_interf}
   \ee
where $\ell$ is the distance from the particle to the plate, here taken to be along the $z$ direction. Let us consider one segment of the path $\Gamma_0$ where the particle evolves between times $t_i$ and $t_j$, where $i,j$ denote the endpoints of the segment.   The associated phase shift is 
\begin{align}
\delta \phi_{ij} & = \int_{t_i}^{t_j} dt \,\, \frac{ m_N\rho }{\Lambda^4}  \frac{1}{4\pi^2}  \int d\q \,\q^2   \frac{e^{ -  2 (z_i+v(t-t_i))\,\gamma_2}}{2 \gamma_2(\gamma_1+\gamma_2)^2}\,
\\& = \frac{1}{v}\frac{ m_N\rho }{\Lambda^4}  \frac{1}{4\pi^2}  \int d\q \,\q^2   \frac{e^{ -  2 z_i\,\gamma_2}-e^{ -  2 z_j\,\gamma_2}}{4  \gamma^2_2(\gamma_1+\gamma_2)^2}\,
\label{eq:phiAB}
\end{align}
This is an exact result following from the exact expression Eq.\,\eqref{eq:V_interf}. If one instead used the approximate expressions for either the Casimir or the Casimir-Polder regime to compute the phase shift, one would need to ensure that all distances involved are respectively  much bigger  or  much smaller than the Compton wavelength in the plate, $m_1^{-1}$ (see section \ref{se:planar}). 
Since in the interferometry experiment the separation between the point source and the plane  varies, using either the Casimir  or the Casimir-Polder approximation   may potentially give an erroneous result. 
The phase shift calculation  provides a concrete example of prediction  for which   the use of  the  exact result Eq.\,\eqref{eq:V_interf} is in general mandatory.

\subsection{Limits}

The phase shift given by Eq.\,\eqref{eq:phiAB} can be further evaluated in some limits when taking $m=0$. All approximations below have been checked numerically.

\subsubsection{ $m_1 z_i\gg 1$, $m_1 z_j\gg 1$}
\label{se:phase_1}

This case amounts to computing the phase shift in the Casimir regime. It is obtained by approximating $\gamma_1+\gamma_2\approx m_1$ in the denominator of Eq.\,\eqref{eq:phiAB}. 
We obtain
\begin{align}
\delta \phi_{ij} & = \int^{t_j}_{t_i} dt \frac{m_N}{32 \pi^2 \Lambda^2 z^2} \\
& = \frac{m_N(z_j-z_i)}{32\pi^2 \Lambda^2\,v \, z_i z_j} \,=
\frac{m_N(t_j-t_i)}{32\pi^2 \Lambda^2 z_i z_j} \,.
\end{align}
The overall factor $\Lambda^{-2}$ is characteristic of the Casimir regime. 

\subsubsection{ $m_1 z_i\ll 1$, $m_1 z_j\ll 1$}
\label{se:phase_2}

This case would amount to computing the phase shift in the Casimir-Polder regime. However, approximating $\gamma_1+\gamma_2\approx 2k$ in the denominator gives a divergent result, therefore we have to go beyond the Casimir-Polder approximation to obtain a finite expression. This is possible only in our formalism:  the small but nonzero effective mass in the plate regularizes the divergence. It is obtained by taking
$\gamma_1+\gamma_2\approx 2k+\frac{m_1^2}{2k}$ in the denominator. 
We obtain
\be
\delta \phi_{ij} = \frac{m_N (t_j-t_i) \rho}
{128 \pi^2  \Lambda^4}\left(
\frac{1}{2}- \gamma + \frac{ z_j \log\left(
\frac{\Lambda}{z_j \sqrt{\rho}}
\right) -
z_i \log\left(
\frac{\Lambda}{z_i \sqrt{\rho}}
\right)
 }{z_j-z_i}
\right)
\ee
The overall $\Lambda^{-4}$ is characteristic of the Casimir-Polder regime. 

\subsubsection{ $m_1 z_i\ll 1$, $m_1 z_j\gg 1$}
\label{se:phase_3}

In this nontrivial case we do not find an accurate approximation, only expressions valid up to $O(1)$ uncertainty, which are nevertheless very useful. 
Taking $\gamma_1+\gamma_2\approx m_1$ in the denominator gives
\be
\delta \phi_{ij} = \frac{m_N (t_j-t_i) \rho}
{16 \pi^2 z_j \Lambda^3}
 \,.
\ee
Taking $\gamma_1+\gamma_2\approx 2k+m_1^2/2k$ in the denominator gives
\be
\delta \phi_{ij} = \frac{m_N (t_j-t_i) \rho}
{128 \pi z_j \Lambda^3} \,.
\ee
The exact result lies in between these two expressions. 
We can see that the overall scaling for this case is $\Lambda^{-3}$.

\subsection{Sensitivity to New Particles}

\begin{figure}
\centering
	\includegraphics[width=1.\linewidth]{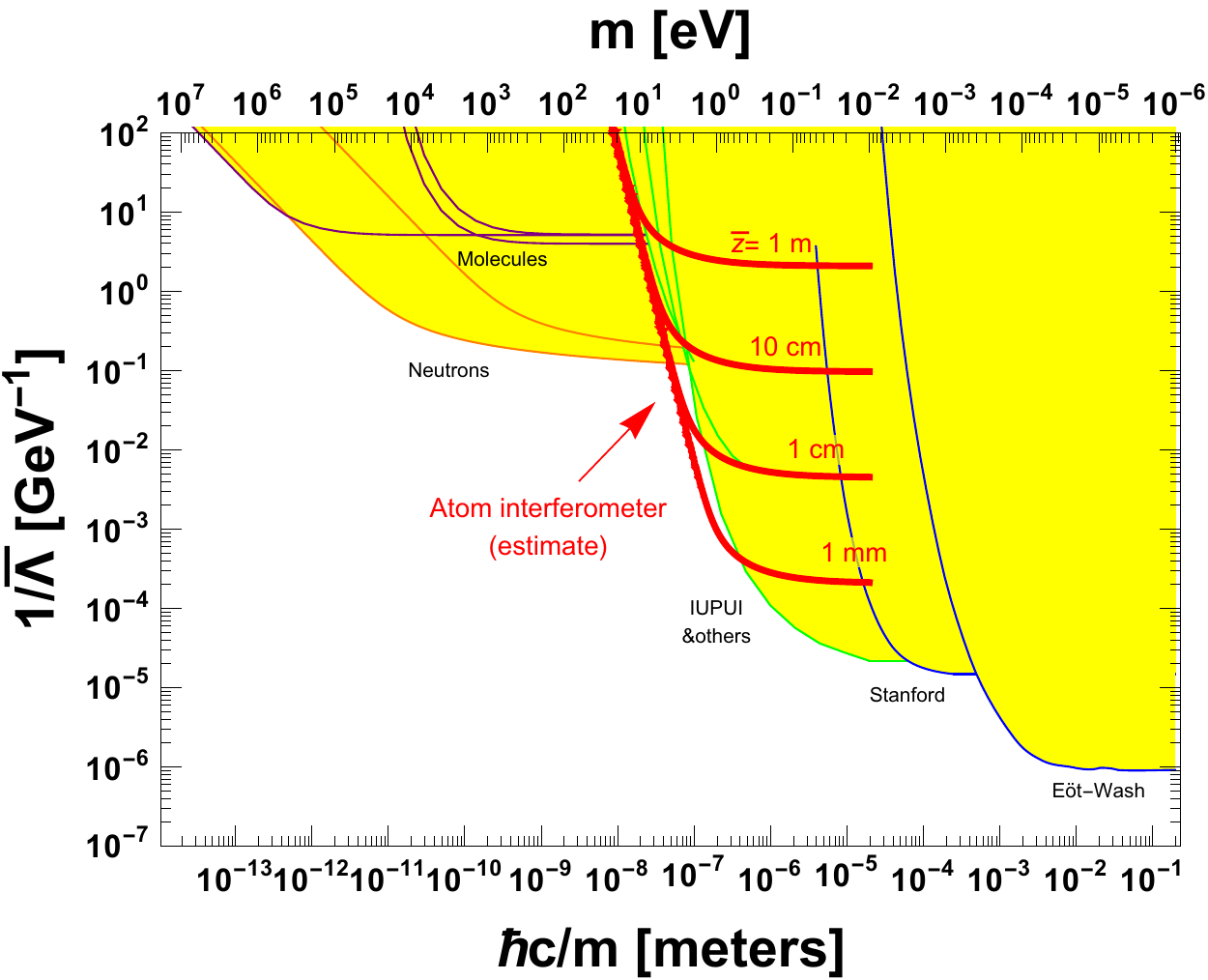}
\caption{ 
Bounds on the quantum force induced by a scalar field bilinearly coupled to nucleons. Red lines correspond to the sensitivity from atom interferometry using $a_{\rm ex}=10^{-8}$m$^2$s$^{-1}$. 
Yellow regions correspond to bounds from other experiments and  match  Ref.\,\cite{Brax:2017xho}. Those were computed in the Casimir-Polder approximation, except the bound from bouncing neutrons.  
\label{fig:landscape}
}
\end{figure}

The phase shift over the closed path $ \Gamma_0$ can be written as $\Delta \Phi = a \kappa T^2 $, where $\kappa= M(v-v')$ is the transferred momentum from the laser pulses and $T$ is the period of pulses  (\textit{i.e.} $2T$ is the total time between splitting ($A$) and recombination ($B$) of the wavepackets, see Fig.\,\ref{fig:diags_interferometry}). The coefficient $a$ has dimension of an acceleration and can be taken as the figure of merit 
for the precision of the atom interferometer. The  sensitivity of existing experiments can typically reach  \be a_{\rm ex}\sim 10^{-9}g =10^{-8}\,{\rm m}^2 {\rm s}^{-1}\ee (see \textit{e.g.} \cite{1999Natur.400..849P}), that we use as our reference value. 

The predicted value of $a$ given by the quantum force in our model is easily obtained in the different regimes discussed above.  
Comparing it to $a_{\rm ex}$ we obtain an experimental bound on the parameters of the $\Phi$ field, \textit{i.e.}  an exclusion region in the $(m,\Lambda)$ plane. For better comparison to other experimental constraints we introduce $\bar \Lambda =  \frac{\Lambda^2}{m_N}$, which then matches the convention in \cite{Brax:2017xho}. 

The result is shown in Fig.\,\ref{fig:landscape}. In the regime relevant for the  presented sensitivities,  the phase shift is dominated by the contributions from the segments near the plate, $\delta\phi\approx \delta\phi_{AC}-\delta\phi_{AD} $. Moreover these contributions are typically in the nontrivial regime of section \ref{se:phase_3}, which depends only on $z_C\approx z_D\equiv \bar z $, which we refer to as  the length of the arm of the interfometer. The sensitivity greatly increases when $\bar z$ decreases, reflecting the fact that the quantum force quickly increases at short distance.  
 The universal diagonal line correspond to the suppression of the quantum force due to the short Compton wavelength of the $\Phi$ field.  A different sensitivity in $a$ would amount to a change in $\bar z$.
 All the lines are in the nontrivial regime of section \ref{se:phase_3}, because for the relevant values of $\Lambda$ the effective Compton wavelength in the plate $m^{-1}_1$ is much smaller than $\bar z$.

Other experimental bounds are shown in Fig.\,\ref{fig:landscape} for comparison.\,\footnote{
{ In the case of dark matter, searches via quantum forces are typically complementary from direct detection searches. The former probe low DM masses while the latter probe high DM masses.  Fig.\,2 in \cite{Fichet:2017bng} illustrates this complementarity. } }
\footnote{ Besides interferometry, another kind of experiment carried out between a particle and a plane
is the neutron bouncer, see \textit{e.g.}
\cite{Nesvizhevsky:2007by, Brax:2011hb,Brax:2013cfa, Jenke:2014yel, Cronenberg:2018qxf, Brax:2017hna}. The quantum levels of the neutrons in the gravitational field of the Earth are probed, via \textit{e.g.}  Rabi oscillation techniques, that constrain  the difference between energy levels. The quantum levels of the neutrons are classified by an integer $n$ and have the energies
$
E_n= m_N g z_0 \epsilon_n
$
where $m_N$ is the neutron's mass, $g$ the acceleration of gravity on Earth and $z_0=(2m_N^2 g)^{-1/3}$ is a characteristic scale of the order of $6\,\mu$m. The number $-\epsilon_n$ are the zeros of the Airy function ${\rm Ai}$ where the wave functions are $\psi_n (z)\propto {\rm Ai}(\frac{z}{z_0}-\epsilon_k)$ and $z$ is the distance to the plate. The perturbation to the energy levels due to the anomalous plate-neutron interaction is given by 
$
\delta E_n= \langle \psi_n\vert V(z) \vert \psi_n\rangle
$
which is constrained by $\vert \delta E_3- \delta E_1\vert \le 10^{-14} {\rm eV}$ \cite{Cronenberg:2018qxf}.  The subsequent bound obtained using  the particle-plane force Eq.\,\eqref{ppo}  is shown in Fig.\,\ref{fig:landscape}. This bound is subdominant with respect to the other bounds --- also when using the Casimir-Polder approximation \cite{Brax:2017hna}.  
}
These other bounds  have been so far computed only in the Casimir-Polder regime. In light of the present work, we can see that, while  this is exact for molecular bounds and neutron scattering,  the Casimir-Polder regime is in general only 
an approximation in the presence of macroscopic bodies.
Using the exact prediction would tend to weaken the bound (see section \ref{se:planar}).


We can thus conclude that atom interferometry turns out to be rather competitive method to search for quantum dark forces, provided the interferometer arms (or at least those near the plate, \textit{i.e.} $AC$ and $AD$) have length below $\bar z \sim 10$\,cm.  This is a reasonable length scale from the experimental viewpoint, which already appears in recent experiments such as the one of \cite{Hamilton:2015zga}.

\section{Conclusion}

\label{se:conc}

How does a classical body  respond to an arbitrary deformation in the quantum vacuum? This question 
can be tackled by  introducing the notion of quantum work $W$,   an observable quantity which goes beyond  the  quantum force $\F$ and reduces to $W= \F \cdot \Lv $ in  specific cases when the deformation flow $\Lv$ is simple enough, \textit{e.g.} when rigid bodies are  displaced with respect to each other. 
In this paper we have studied the quantum work induced by a massive scalar field bilinearly coupled to macroscopic bodies made of classical matter.
Unlike for abstract sources, the number densities of such bodies must satisfy the local conservation of matter. We have shown that the prediction of the quantum work turns out to be finite --- up to physical,  renormalizable divergences --- upon requesting conservation of matter.   This result applies to any shape and geometry, either rigid or deformable. 
This is shown both for a renormalizable --- possibly strongly-coupled --- theory, and in a more general effective field theory setup allowing for higher derivative interactions between the scalar and matter. 

Our result about finiteness of the quantum work readily explains why the QFT prediction of quantum forces sometimes   feature seemingly ``unremovable" divergences for certain  geometries. A key example is the  quantum work felt by a Dirichlet sphere under a radial deformation. The radial deformation flow is not divergence-free and thus the sphere density must vary to ensure matter conservation. Not taking this into account implies that matter of the sphere is not conserved, thus the finiteness property is not ensured, and therefore  the expression of the quantum force can have a spurious divergence. We have  explicitly verified that taking into account the  variation of the density removes the spurious  divergence in the case of the Dirichlet sphere. 

When specializing  to rigid bodies, the quantum work leads to a quantum force that reduces to  the scalar Casimir and Casimir-Polder forces as special limits.   There is a clear diagrammatic understanding of  this interpolation. In the short distance regime,  the main contribution comes from the loop with only one coupling to each body, which corresponds to Casimir-Polder. In the long distance regime, loops with arbitrary number of insertions contribute, but their resummation amounts to having a Dirichlet condition on the boundary of the bodies, which corresponds to the traditional Casimir setup.

We have computed the quantum forces in plate-plate and plate-point geometries. If, for example, the scalar has zero fundamental mass, in the plate-plate geometry the force behaves as $\frac{1}{\ell^4}$ at long distance (Casimir) and as  $\frac{1}{\ell^2}$ at short distance (Casimir-Polder). For plate-point geometry the force behaves as $\frac{1}{\ell^3}$ at long distance (Casimir) and as  $\frac{1}{\ell}$ at short distance (Casimir-Polder). For nonzero fundamental mass the two limits can still exist, but the Casimir force drops exponentially at separation  $\ell \gg \frac{1}{2m}$. 

These results have concrete applications for physical setups aimed at searching for light dark  particles. Such particles are ubiquitous in dark matter models, dark energy models, and in extensions of the Standard Model.  
Here  we have briefly illustrated an application  of the point-plane quantum force 
 for atom interferometers. In this type of experiment, a non-relativistic atom with trajectory in the vicinity of a large sphere or a plane is sensitive to the existence of a new force, which induces a phase shift that can be computed in the WKB approximation.
  We have computed the atomic phase shift induced by the scalar quantum force, and point out that the full result (as opposed to a simple Casimir or Casimir-Polder approximation) is required to  obtain a sensible prediction. 
 Using inputs from existing experiments, we show that atom interferometry is likely to become a competitive method to search for light dark particles bilinearly coupled to matter, provided that the interferometer arms are shorter than $\sim 10$\,cm, as summarized in Fig.\,\ref{fig:landscape}.  
 This is a reasonable length scale from the experimental viewpoint,  which already appears in setups such as the one of \cite{Hamilton:2015zga}.

In  future work we will further investigate/revisit the constraints and sensitivities of future experiments to macroscopic dark quantum forces with Casimir-like behaviors.

\section*{Acknowledgments}

SF thanks Daniel Davies for a useful discussion. This work has been supported by the S\~ao Paulo Research Foundation (FAPESP) under grants \#2011/11973, \#2014/21477-2 and \#2018/11721-4, by CAPES under grant \#88887.194785, and by the University of California, Riverside.

\appendix

\section{Derivation of the Plate-Point Casimir-Polder Potential}
\label{app:CP}

The Casimir-Polder potential between a particle and a plate can be obtained by direct calculation in the weak coupling regime (\textit{i.e.} the limit of small density in the plate).  
We start by  computing the potential between two point sources. The corresponding source term is
\be
{\cal L} \supset -\frac{1}{2} \Phi^2
\left(
 \frac{m_a}{\Lambda^2} \delta^{3}(\x-\x_a) +   \frac{m_N}{\Lambda^2} \delta^{3}(\x-\x_b) \right)\,.
\ee
We remind that this is the non-relativistic approximation of the 4-point interaction
\be
{\cal L} \supset -\frac{1}{2} \Phi^2
\left(
 \frac{m_a}{\Lambda^2} \bar\psi_a \psi_a +   \frac{m_N}{\Lambda^2} \bar\psi_b \psi_b \right)\,
\ee
between the scalar and two fermion species. 
The scattering amplitude with two insertions of two particle-anti particles pairs leads to a  bubble diagram  which reads
\be
i{\cal M} = - \frac{m_N m_a}{\Lambda^4} \,4 m_N m_a\, \frac{1}{2} \int \frac{d^3k}{(2\pi)^3} \frac{e^{i\omega_2|z_1-z_2|)}}{2\omega_2}
\frac{e^{i\omega'_2|z_1-z_2|)}}{2\omega'_2}
\ee
where $\omega_2=\sqrt{k^2-m_2^2}$, $\omega'_2=\sqrt{(k+p)^2-m_2^2}$. The factor of $\frac{1}{2}$ is a symmetry factor and the external fermions are such that their nonrelativistic wavefunctions are normalised as $\bar u_a u_a= 2 m_a, \ \bar u_b u_b= 2m_N$.
The non-relativistic  scattering potential is given by
\begin{eqnarray}
&&\tilde V(p,z_1-z_2) = - \frac{\cal{M}}{4  m_a m_N}\nonumber \\
&&= -i \frac{m_N m_a}{\Lambda^4} \frac{1}{2}  \int \frac{d^3k}{(2\pi)^3} \frac{e^{i\omega_2|z_1-z_2|)}}{2\omega_2}
\frac{e^{i\omega_2'|z_1-z_2|)}}{2\omega_2'}\,.\nonumber \\
\end{eqnarray}
The spatial potential is  then obtained from the Fourier transform of $\tilde V$,
\be
V\left(\sqrt{(z_1-z_2)^2+\x_\parallel^2}\right)= \int \frac{d^2 {\bm p}_{\parallel} }{(2\pi)^2} \tilde V({\bm p}_{\parallel},z_1,z_2)e^{i {\bm p}_{\parallel}\cdot \x_{\parallel} }
\ee
where $\x_\parallel=(x_1,x_2 )$.

We now consider an ensemble of $N_1$ particle of the species $a$ in a volume $V_1$ with a number density $n_1=\frac{N_1}{V_1}$. 
We  average the potential over the plate with a separation $\ell$ to the point particle as
\be
V(\ell)=n_1\int d^2\x_{\parallel}\int_{-\infty}^0 dz_1
 \int \frac{d^2 {\bm p}_{\parallel} }{(2\pi)^2} \tilde V({\bm p}_{\parallel},z_1,\ell)e^{i {\bm p}_{\parallel} \cdot \x_{\parallel} }.
\ee
The transverse integrals simplify and the potential  becomes simply 
\be
V(\ell)=  n_1 \int_{-\infty}^0 dz_1    \tilde V(0,z_1,\ell) 
 = -  n_1 \frac{m_a m_N}{\Lambda^4}
 \, \int \frac{d^3k_E}{(2\pi)^3}
 \frac{e^{-2\gamma_2 \ell }}{16(\gamma_2)^3}
\ee
after Wick's rotation. Using $m_1^2-m_2^2= n_1 m_a/\Lambda^2$, we find
\be 
V(\ell)= -  (m_1^2-m_2^2 )  \frac{m_N}{\Lambda^2}
 \, \int \frac{d^3k_E}{(2\pi)^3}
 \frac{e^{-2\gamma_2 \ell }}{16(\gamma_2)^3}\,.
\ee
Finally the force is obtained by taking the derivative with respect to $\ell$
\be
F=-\partial_\ell V=-(m_1^2-m_2^2 ) \frac{m_N}{\Lambda^2}
 \, \int \frac{d^3k_E}{(2\pi)^3}
 \frac{e^{-2\gamma_2 \ell }}{8(\gamma_2)^2}\,.
\ee
This  reproduces  Eq.\,\eqref{eq:CP}.

\section{On Divergences from the Heat Kernel Expansion}
\label{app:HK}

In this appendix we review how the heat kernel expansion allows for a  proof of the finiteness of the quantum work at one-loop level under the assumptions that \textit{(i)} the bodies are incompressible and \textit{(ii)} the fluctuation is  repelled from the sources so that we can set boundary conditions at the source surfaces, analogously to the Dirichlet limit described in section \ref{se:casimir_limit}.
We will then see that the argument fails when condition \textit{(i)} is dropped \textit{i.e.} considering a compressible body, which signals the need for an approach that includes matter conservation.

We essentially review  the exposition from \cite{bordag2009advances}.
 We consider two rigid bodies such that $J=J_1+J_2$ as in section \ref{se:Casimir}. 
First of all, let us rewrite the scalar action as
\be
S(\Phi)= S_2(\Phi) + S_{\rm int}(\Phi)
\ee
where we have included the  $J$-dependent contribution to the effective mass in the quadratic part of the action.
This action reads 
\be  
S_2(\Phi)= -\frac{1}{2}\int d^4 x \Phi \Box_J \Phi
\ee
where $\Box_J$ is the operator which includes the effective mass term. In the renormalizable case, this is simply
\be
\Box_J= \Box + m^2 + \frac{J}{\Lambda}\,.
\ee
 In the EFT case, higher order derivatives are present. 
 
We assume weak coupling.  When the sources are static, the quantum vacuum energy  reads
\be  
E(J)= \frac{i}{T}\ln Z(J)= -\frac{i}{2T}{\rm{Tr}}\ln \Box_J + \frac{i}{T}W_2(J)
\ee
where the fist contribution is the one-loop contribution to the vacuum diagrams, and $W_2(J)$ is the sum over all connected vacuum diagrams at two loops and higher. 
Here we focus on the one-loop contribution, which is independent of other possible interactions ( \textit{e.g.}  polynomial self-interactions of $\Phi$). The one-loop piece can be evaluated using the heat kernel $K_J= e^{-t \Box_J}$ and its trace as
\be  
{\rm Tr}\ln \Delta_J= -\int_0^\infty \frac{dt}{t} {\rm Tr}\,K_J(t)\,.
\ee
See \cite{Vassilevich:2003xt} for a conceptual review on heat kernel methods. 
The trace of the heat kernel has a $t\to 0$ expansion
\be
{\rm Tr}\,K_J(t)= \frac{1}{(2\pi t)^{3/2}}(a_0+ a_{1/2}t^{1/2}+ a_1 t + a_{3/2}t^{3/2}+\dots)
\ee
We can see that the heat kernel coefficients $a_{n/2},\ n=0,\dots 3$ are responsible for the divergences of $E(J)$. These coefficients are universal local quantities \cite{Vassilevich:2003xt,bordag2009advances} and can be expressed as volume integrals over the support of the $J_i$ sources and surface integrals over their boundaries, $\partial J_i$. 
The first coefficient simply amounts to the volume  of the support of the fluctuation, which is the space outside of the sources,
\be
a_0=V-V_1-V_2
\ee
with $V_{1,2}={\rm Vol}(J_{1,2})$ and $V$ the volume of the total space.   Depending on the geometric setup, these various volumes  may be finite or infinite, this is a minor detail in the argument.   

We can now vary the quantum vacuum energy under some deformation flow. We use the notation of section \ref{se:quantumwork}.  
In the case of a rigid deformation flow (\textit{i.e.} satisfying $\bpartial \cdot \Lv =0$),  \textit{e.g.} two bodies moving apart, the coefficients are independent of the relative position of the bodies \cite{Fulling:2003zx, bordag2009advances}. 
As a result all the divergences cancel when computing the quantum work by varying the relative position of the bodies. This is easily illustrated at the level of the $a_0$ coefficient, for which  we simply have
\be
\partial_\lambda a_0=- \partial_\lambda\int_{J_{\lambda}} d^d {\x}= - \int_{J_{\lambda}}  d^d {\x} \,\bpartial\cdot \Lv  =0\,.
\ee
where we have used that $\partial_\lambda V=0$. Therefore under the assumption of {{incompressibility}} the quantum work is finite. This is a version of the classic argument given in \cite{bordag2009advances}. 

In contrast, in the case of a deformation  flow for which $\bpartial\cdot \Lv\neq 0 $, \textit{i.e.} dropping the assumption of {{incompressibility}}, we can see that $\partial_\lambda a_0 \neq 0 $, which then results in a divergence in the quantum work. There is no  parameter in the Lagrangian  into which this divergence could be absorbed, therefore there is no hope to renormalize this divergence away. The only way out we found is to let the number density of the sources  vary in such way that matter conservation in the source is satisfied. As shown in section \ref{se:quantumwork}, \ref{se:Wfinite_EFT} and exemplified in section~\ref{se:Sphere}, this is the  condition that  ensures that all possible divergences in the quantum work vanish.

\section{The Loop Divergence from Momentum Space}

\label{se:loop_divergence}


We are interested in the loop integral given in Eq.\,\eqref{eq:loop_points}, 
\be 
I=\left(\prod^{q}_{i=1} \int d^d\bmu_i J(\bmu_i) \int dt_i \right)  \prod^{q}_{i=0} {\cal B}''\Delta_0(\mu_i,\mu_{i+1})\bigg|_{\mu_0=x, \mu_{q+1}=\xeps}.
\ee
We  introduce  the $(d+1)$-dimensional Fourier transform of the propagators  defined as
\be 
\Delta_0(\mu_i,\mu_{i+1})= \int \slashed{d}^{d+1} { k} e^{i{k.(\mu_{i+1}-\mu_i})}\tilde \Delta_0 ( k)\,.
\ee
We introduce the $d$-dimensional Fourier transform of the sources, 
\be
J(\bmu_i)=  \int \slashed{d}^{d} {\bf k} e^{i {\bf k} \cdot  \bmu_i} \tilde J({\bf k})\,.
\ee
We have defined  $\slashed{d}k= \frac{dk}{2\pi}$. The contraction between the $(d+1)$-vectors is done using the Minkowski metric $k\cdot y= \eta_{\mu\nu} k^{\mu} y^\nu$ with $k^\mu=(\omega,\bf k )$. 
The loop integral becomes
\be 
I= \int \slashed{d}^{d+1}{ k}_0 \left(  \prod_{i=1}^{q} \int  \slashed {d}^d {\bf p}_i   \tilde J({\bf p}_i) \right) \left(\prod_{i=0}^{q} \tilde B''(k_{i-1},k_i)  \tilde \Delta_0 (k_i) \right) e^{ik_{q}\cdot x_\epsilon-i k_0 x}
\ee
with ${\bf p}_i={\bf k}_{i}-{\bf k}_{i-1}$ the momentum transferred to the source. 

Performing the time and position integrals we obtain an integral over the  $(d+1)$-momentum $k_0$ running through the loop.  
 The derivative operators in real space become simply multiplying operators depending on the two momenta entering each vertex $\tilde B''(k_{i-1},k_i)$. In the EFT case it contains  at least two powers of  the momenta.
 
Since the source $J$ has compact support, the Riemann-Lebesgue lemma guarantees that $\tilde J({\bf p})$ vanishes for large momenta and in fact falls off for $\vert {\bf p}\vert \gtrsim R_J^{-1}$. 
This  implies  a  smooth cutoff on the value of ${\bf p}$. 
On the other hand the divergence in $I$ occurs for arbitrarily large loop momentum $k_0$ . Since the $\bf p$ are bounded, at large loop momentum we have $k_0\simeq k_1  \simeq \ldots \simeq k_q$,  thus the divergent part of the diagram is $\bf p$-independent and is only a function of $({ x}_\epsilon-{x})=\epsilon$. Writing $c_J=\int  \slashed {d}^d {\bf p}_i   \tilde J({\bf p}_i) $, we have the structure
\be
I\approx c_J^q  L^{\rm div}_\epsilon \,
\ee
where $L^{\rm div}_\epsilon$  is the position-independent  divergent quantity. 

\bibliographystyle{JHEP}
\bibliography{biblio}

\end{document}